\documentclass[twocolumn,showpacs,preprintnumbers,amsmath,amssymb,prb,superscriptaddress]{revtex4}
\usepackage{mathrsfs}
\usepackage{graphicx}
\usepackage{dcolumn}
\usepackage{bm}
\usepackage{amsmath}
\usepackage{amsfonts}
\usepackage{color}
\usepackage{hyperref}

\begin{document}

\title{Chern-Simons Theory and Wilson Loops in the Brillouin Zone}
\author{Biao Lian}
\affiliation{Department of Physics, McCullough Building, Stanford University, Stanford, California 94305-4045, USA}
\author{Cumrun Vafa}
\affiliation{Jefferson Physical Laboratory, Harvard University, Cambridge, MA 02138, USA}
\author{Farzan Vafa}
\affiliation{Department of Physics, University of California, Santa Barbara, CA 93106, USA}
\author{Shou-Cheng Zhang}
\affiliation{Department of Physics, McCullough Building, Stanford University, Stanford, California 94305-4045, USA}


\begin{abstract}
Berry connection is conventionally defined as a static gauge field in the Brillouin zone.  Here we show that for three-dimensional (3d) time-reversal invariant superconductors, a generalized Berry gauge field behaves as a fluctuating field of a Chern-Simons gauge theory.  The gapless nodal lines in the momentum space play the role of Wilson loop observables, while their linking and knot invariants modify the gravitational theta angle.  This angle induces a topological gravitomagnetoelectric effect where a temperature gradient induces a rotational energy flow.  We also show how topological strings may be realized in the 6 dimensional phase space, where the physical space defects play the role of topological D-branes.
\end{abstract}

\date{\today}


\maketitle

\section{Introduction}

Topology has played an increasingly important role in condensed matter physics.  An early application involved using topological Chern-Simons theory in quantum hall systems \cite{Zhang1989,Blok1990,Zhang1992,Moore1991,Lu2012}.  More recently, topological ideas applied to Brillouin zone (BZ) have played a significant role in classifying topological phases of matter (for example, see \cite{Qi2008,Qi2011,Xiao2010} and references therein). Whereas the former application involved a dynamical gauge field in physical spacetime, the latter cases were formulated in the momentum space and involved topological aspects of static gauge fields such as the Berry connection.  It is thus natural to ask: Can dynamical or fluctuational gauge fields naturally occur in a condensed matter system?
The aim of this paper is to argue that this can be done at least in the context of three-dimensional (3d) time reversal invariant (TRI) superconductors, where (a slightly modified version of) Berry's gauge field can be viewed as fluctuational in the Brillouin zone and governed by the Chern-Simons field theory. In this class of theories, the fluctuations of the Berry connection are induced from quantum fluctuations of the superconductor's pairing amplitude. More interestingly, 3d TRI superconductors are known to be capable of having gapless nodal lines in the BZ \cite{Yip2014,Volovik1993}, and we will show that they play exactly the role of Wilson loop observables in the Chern-Simons theory.

In nature, a large class of materials belongs to 3d TRI superconductors, which includes most of the conventional s-wave superconductors described by the Bardeen-Cooper-Schrieffer theory, and a large number of unconventional superconductors such as many heavy fermion superconductors, cuprates, iron-based superconductors and the 3d topological superconductors. In particular, unconventional superconductors often exhibit a highly anisotropic pairing amplitude such as p-wave, d-wave or their hybridizations with s-wave, and have much richer phenomena in experiments \cite{Yip2014,Volovik1993,Lee1993,Graf1996}. Previous studies have shown from different perspectives the significance of topology in these TRI unconventional superconductors. Gapped 3d TRI superconductors are shown to have a $\mathbb{Z}$ topological classification \cite{Qi2011,Qi2010t,Kitaev2009,Ryu2010}, which fit in the generic K-theory classification framework of gapped topological phases \cite{Kitaev2009,Ryu2010}. Topological superconductors are defined as such superconductors with a nonzero topological number (e.g. the $^3$He-B phase), and are shown to support gapless topological Majorana fermions on the surfaces \cite{Qi2011}.
On the other hand, the gapless 3d TRI superconductors contain gapless nodal lines in the BZ that are allowed by the time-reversal symmetry, with known examples such as the heavy fermion superconductor CePt$_3$Si and the cuprates \cite{Yip2014,Volovik1993}. Several recent studies show that topological numbers can also be defined for these nodal-line superconductors in terms of K-theory \cite{Horava2005,Zhao2013,Chiu2014}, which give rise to various types of topological Majorana surface states \cite{Schnyder2012,Shunji2013,Schnyder2015}.
In addition, in analogy to 3d Weyl semimetals which can be viewed as intermediate phases between 3d topological insulators and trivial insulators \cite{Haldane2004,Murakami2007}, nodal-line superconductors can also arise as gapless intermediate phases between conventional TRI and topological TRI superconductor phases \cite{Hosur2014}. These different while related facts strongly indicate the existence of a unified topological field theory that describes both gapped and gapless 3d TRI topological superconductors. This also motivates us to consider the Chern-Simons theory in the 3d BZ for these superconductors.

There is already a hint that a topological field theory in BZ can be physically relevant.  In particular, it has been shown \cite{Qi2008,Li2010,Bulmash2015} that
for an insulator in odd spatial dimensions, the value of the Chern-Simons (CS) action for Berry connection of the filled bands in the BZ  computes the effective theta angle of the corresponding $c_n(F)=F^n$ term (or a gravitational analog proportional to $R\wedge R$ in the 3d case \cite{Wang2011,Furusaki2013}), where $F$ is the electromagnetic field in the physical spacetime.  In other words, there is a coupling of the form
\begin{equation}\label{level}
S\propto \left[ \int_{T_{BZ}^{2n-1}} \text{CS}(a_\text{Berry}) \right]\times \left[\int_{\mathbb R^{2n}} F^n\right]\ .
\end{equation}%
We will be specializing to the case of $n=2$, i.e., 3d space in this paper.
(A similar term can be used to compute the gravitational response involving $\int R \wedge R$).
Note that the CS action, which is an angular quantity and has a shift ambiguity, has the correct structure to be the coefficient in front of
$F\wedge F$, which needs to be defined only up to shift symmetry.  In this context, it is very natural to ask whether the Berry connection $a_\text{Berry}$ can fluctuate. In particular, we can imagine having in physical spacetime a pulse where $(1/8\pi^2) \int F\wedge F=k$, which leads to
an effective level of $k$ for the CS theory in the BZ. Can the Berry connection behave as if it obey the CS theory?  For the answer to be yes, the classical background for
$a_\text{Berry}$ must be flat, as is demanded by the CS equations of motion.  This is certainly not the case in general.  However,
as we shall see, it is the case for TRI superconductors with a slightly modified Berry connection.
With this encouraging result, one may then ask whether there are natural objects in the BZ corresponding to Wilson loops of the CS theory.
Indeed, for TRI superconductors the symmetries allow gapless nodal lines in the BZ, which we will see end up playing the role of Wilson lines for the CS theory.
For superconductors, the relevant term to compute is the gravitational $\int R\wedge R$ term.  But with the gapless modes, as would be the case if we have nodal lines, the theta angle is ambiguous.  It turns out the choices of resolving this ambiguity by introducing infinitesimal time reversal breaking perturbations to get rid of gapless modes are in 1-1 correspondence with allowed basic charges for the CS theory!
We thus find that dressing up the nodal lines with this data gives an unambiguous theta angle for the coefficient of $R\wedge R$ term, which is identified with the free energy of CS theory in the presence of Wilson loops and leads to physically measurable effects. In particular, as we shall see, when the nodal lines change from linked to unlinked, the theta angle changes in units of $\pi$.
Our main discussion is in the context of the abelian Chern-Simons theory, but we also indicate briefly how the theory gets extended to non-abelian case, and how in particular the $U(2)$ case can be potentially realized in experiments.
We will also connect aspects of our discussions with topological strings \cite{Witten:1992fb} which is formulated in a 6-dimensional symplectic manifold (which is typically taken to be a Calabi-Yau 3-fold).  In our context, the phase space which is $T^*T^3$ turns out to play the role of this 6d symplectic space.  In the topological string setup, CS theory lives on 3d Lagrangian defects (`D-branes').  If they are oriented along the BZ $T^3$, they give rise to the CS theory we find in the BZ.  In this context, Wilson loop observables arise from a pair of such 3d defects intersecting along a loop \cite{Ooguri2000}.   We will show that line defects in physical space lead to such 3d Lagrangian branes and can also give rise to Wilson loop observables (nodal lines) in the BZ.

The organization of this paper is as follows. In section II, we review the Hamiltonian of TRI superconductors and define the modified Berry connection. In section III, we relate the topological $\theta$ angle to a Chern-Simons term. Section IV incorporates the nodal lines as Wilson loops, first in minimal model and then in multi-band system. In section V, we consider non-abelian nodal lines, first in a $U(2)$ example and then generalize to the $U(N)$ example. In Section VI we  discuss connections with topological strings and show how line defects in physical space also lead to effective Wilson loops in the BZ. We present our conclusions in section VII, and in the appendix are some computational details.

\section{Hamiltonian and the Berry connection}

At the single-particle level, superconductors are described by the Bogoliubov-de Gennes (BdG) Hamiltonian which has an inherent charge conjugation symmetry $\mathbf{C}$. For a superconductor with $N$ electron bands, the BdG Hamiltonian can be written in terms of the Nambu basis $\Psi_\mathbf{k}=(\psi_\mathbf{k}^T, \psi_{-\mathbf{k}}^\dag)^T$ as $H=\frac{1}{2}\sum_\mathbf{k}\Psi_\mathbf{k}^\dag H(\mathbf{k})\Psi_\mathbf{k}$, where
\begin{equation}
H(\mathbf{k})=\left(\begin{array}{cc}h(\mathbf{k})&\Delta(\mathbf{k})\\ \Delta^\dag(\mathbf{k})&-h^T(-\mathbf{k})\end{array}\right)
\end{equation}
is a $2N\times2N$ matrix, $\mathbf{k}$ is the momentum, and $\psi_\mathbf{k}=(\psi_{1,\mathbf{k}},\cdots, \psi_{N,\mathbf{k}})^T$ is the $N$-component electron basis of the system. Both $h(\mathbf{k})$ and $\Delta(\mathbf{k})$ are $N\times N$ matrices. $h(\mathbf{k})$ represents the single-particle Hamiltonian of the system before superconductivity arises, while $\Delta(\mathbf{k})$ is the pairing amplitude satisfying $\Delta(\mathbf{k})=-\Delta^T(-\mathbf{k})$ as required by the fermion statistics. The charge conjugation is defined as $\mathbf{C}^{-1}\psi_\mathbf{k}\mathbf{C}=\psi^{\dag T}_{-\mathbf{k}}$, or equivalently $\mathbf{C}^{-1}\Psi_\mathbf{k}\mathbf{C}=C_S\Psi_\mathbf{k}$, where $C_S=\tau_1\otimes I_N$ is a $2N\times2N$ matrix, with $\tau_{1,2,3}$ denoting the Pauli matrices (for particle-hole basis) and $I_N$ the $N\times N$ identity matrix. The charge conjugation symmetry of the BdG Hamiltonian can be seen via the relation $C_S^\dag H(\mathbf{k})C_S=-H^T(-\mathbf{k})$.

TRI superconductors are a large class of superconductors that are quite robust in nature. In the presence of time reversal symmetry $\mathbf{T}$, electrons which are time-reversal Kramers pairs have the maximal spatial wave-function overlap, thus fall into cooper pairs most easily \cite{Anderson1959}. For fermions, the time reversal symmetry $\mathbf{T}$ is anti-unitary and satisfies $\mathbf{T}^2=-1$. It acts on the fermions as \cite{Qi2010t,trs} $\mathbf{T}^{-1}\psi_{\mathbf{k}}\mathbf{T}=\mathcal{T}\psi_{-\mathbf{k}}$, where $\mathcal{T}$ is an $N\times N$ matrix satisfying $\mathcal{T}\mathcal{T}^*=-I_N$, $\mathcal{T}^\dag\mathcal{T}=I_N$. The BdG Hamiltonian of a TRI superconductor can then be shown to satisfy $\mathcal{T}^\dag h^T(\mathbf{k})\mathcal{T}=h(-\mathbf{k})$ and $\Delta(\mathbf{k})\mathcal{T}=\mathcal{T}^\dag\Delta^\dag(\mathbf{k})$, namely, $\Delta(\mathbf{k})\mathcal{T}$ is Hermitian. These conditions can be more compactly written as $T_S^\dag H^T(\mathbf{k})T_S=H(-\mathbf{k})$, where $T_{S}=\mbox{diag}(\mathcal{T},-\mathcal{T}^\dag)$ is the $2N\times2N$ time-reversal transformation matrix of the Nambu basis.

It is useful to define the chiral transformation $\chi=iC_ST_S$, which is a unitary Hermitian matrix that anti-commutes with the BdG Hamiltonian, $\chi H(\mathbf{k})=-H(\mathbf{k})\chi$. Upon diagonalizing $\chi$ to $\widetilde{\chi}=\tau_3\otimes I_N$ under a new basis $\widetilde{\Psi}_\mathbf{k}=(\psi_\mathbf{k}^T+i\psi_{-\mathbf{k}}^\dag\mathcal{T}^\dag, \psi_\mathbf{k}^T-i\psi_{-\mathbf{k}}^\dag\mathcal{T}^\dag)^T/\sqrt{2}$, the BdG Hamiltonian becomes
\begin{equation}\label{Jk}
\widetilde{H}(\mathbf{k})=\left(\begin{array}{cc}0&J(\mathbf{k})\\ J^\dag(\mathbf{k})&0\end{array}\right)\ ,
\end{equation}
where $J(\mathbf{k})=h(\mathbf{k})-i\Delta(\mathbf{k})\mathcal{T}$. Since both $h(\mathbf{k})$ and $\Delta(\mathbf{k})\mathcal{T}$ are Hermitian, $J(\mathbf{k})$ is a general $N\times N$ complex matrix. If $J(\mathbf{k})$ is non-singular everywhere, the superconductor is in a fully gapped phase. On the other hand, the superconductor becomes gapless (nodal) at momentum $\mathbf{k}$ if $\det (J(\mathbf{k}))=0$. In the absence of additional symmetries other than the time reversal symmetry, the gapless submanifolds in the momentum space are nodal points for two-dimensional (2d) superconductors, and are 1d nodal lines for 3d superconductors \cite{Zhao2013,Shunji2013,Chiu2014}. When the superconductor is centrosymmetric, the nodal lines (points) in 3d (2d) become doubly degenerate, as we shall show in Sec. \ref{SecU2}. In this paper we focus on 3d superconductors with nodal lines, which are widely found in noncentrosymmetric superconductors \cite{Yip2014} and quasi-2d centrosymmetric superconductors such as cuprates and iron-based superconductors \cite{Tsuei2000,Wangf2011}.

In general, the matrix $J(\mathbf{k})$ can be singular-value-decomposed (SVD) into $J(\mathbf{k})=U_\mathbf{k}^\dag D_\mathbf{k}V_\mathbf{k}$, where $U_\mathbf{k}$, $V_\mathbf{k}$ are unitary matrices, and $D_\mathbf{k}$ is a diagonal matrix with all elements real and nonnegative. The BdG Hamiltonian can then be diagonalized as
\begin{equation}\label{Hdiag}
\Lambda^\dag_\mathbf{k}\widetilde{H}(\mathbf{k})\Lambda_\mathbf{k}=\left(\begin{array}{cc}D_\mathbf{k}&\\&-D_\mathbf{k}\end{array}\right),\  \Lambda_\mathbf{k}=\frac{1}{\sqrt{2}}\left(\begin{array}{cc}U^\dag_\mathbf{k}&U^\dag_\mathbf{k}\\-V^\dag_\mathbf{k}&V^\dag_\mathbf{k}\end{array}\right).
\end{equation}

In literature \cite{Berry1984,Qi2008,Qi2011,Xiao2010}, it is conventional to define a $1$-form U($N$) Berry connection for the $N$ occupied bands with negative energy as $a'_{\alpha\beta}(\mathbf{k})=i\langle\alpha,\mathbf{k}|d|\beta,\mathbf{k}\rangle$, where $d=dk^i\partial_{k^i}$ is the exterior derivative in the momentum space, and $|\alpha,\mathbf{k}\rangle$ is the eigenstate wave function of the $\alpha$-th occupied band at momentum $\mathbf{k}$. This can be rewritten into matrix form as
\begin{equation}\label{aprime}
a'(\mathbf{k})=\frac{i}{2}(UdU^\dag+VdV^\dag)\ ,
\end{equation}
where we have used $U$, $V$ short for $U_\mathbf{k}$, $V_\mathbf{k}$. Berry connection $a'(\mathbf{k})$ defined in this way proves to be important in characterizing the topology of non-interacting gapped systems \cite{Qi2008,Qi2011,Xiao2010}.
However, one should note that the U($N$) gauge freedom of $a'(\mathbf{k})$ is not a symmetry of the Hamiltonian, since the eigenstates $|\alpha,\mathbf{k}\rangle$, if not degenerate, are generically fixed by the Hamiltonian up to a phase factor (see Appendix \ref{AppBerry}). Instead, the U($N$) gauge freedom of $a'(\mathbf{k})$ should be thought of as a symmetry of the projection operator onto the set of occupied bands \cite{Qi2008}.

For the purpose of this paper, we wish to find a gauge field in the BZ whose U($N$) gauge symmetry is a symmetry of the Hamiltonian, so that the CS theory of such a gauge field can serve as an effective theory of the system. For superconductors, this can be achieved by defining a modified U($N$) Berry connection as $a_{mn}(\mathbf{k})=i\sum_\alpha\langle0|\psi_{m,\mathbf{k}}|\alpha,\mathbf{k}\rangle d\langle \alpha,\mathbf{k}|\psi_{n,\mathbf{k}}^\dag|0\rangle$, where $|0\rangle$ is the vacuum state of no electrons, and $\alpha$ runs over all the $2N$ eigenstates. The physical meaning of $a(\mathbf{k})$ from the definition is the Berry connection felt by an electron when its wave function is projected onto the BdG eigenstates and evolves adiabatically. In the TRI case where the Hamiltonian takes the form of Eq. (\ref{Hdiag}), such a Berry connection can be rewritten as a form slightly different from $a'(\mathbf{k})$ in Eq. (\ref{aprime}):
\begin{equation}\label{a}
a(\mathbf{k})=\frac{i}{2}(U^\dag dU+V^\dag dV)\ .
\end{equation}
Alternatively, one can view $a$ as the standard Berry connection for eigenstates of an associated Hamiltonian with $J(k)=U_k D_k V^\dagger _k$, i.e., with $U_k\rightarrow U_k^\dagger$ and $V_k\rightarrow V_k^\dagger$.
This definition of $a$ is closely related to the conventional Berry connection $a'$, as we shall see in Sec. \ref{SecCSA}. 
In particular, in the case $N=1$, we have $a=-a'$, and the two definitions are identical up to a sign.
In the generic $N$ band case, although the eigenstates are fixed by the Hamiltonian, the U($N$) gauge transformation of $a$ simply corresponds to a momentum-dependent unitary transformation of the electron basis $\psi_\mathbf{k}\rightarrow g(\mathbf{k})\psi_\mathbf{k}$, where $g(\mathbf{k})$ is a U($N$) matrix, therefore is a gauge symmetry of the system. By the definition, the new Berry connection $a$ transforms as $a\rightarrow g^\dag ag+ig^\dag dg$, which is exactly a U($N$) gauge transformation. This unitary transformation, however, does not yield a gauge transformation of the conventional Berry connection $a'$(see Appendix \ref{AppBerry}). In this sense, the modified Berry connection $a$ is a U($N$) gauge field that is more natural for the purpose here. Accordingly, a slow external potential $V_{\text{ext}}(\mathbf{x})$ added to the system will induce an effective Hamiltonian perturbation $\delta H=\sum_\mathbf{k} \psi^\dag_\mathbf{k}V_{\text{ext}}(i\nabla_\mathbf{k}+\bm{a})\psi_\mathbf{k}$ in the momentum space.


\section{Topological theta angle as a Chern-Simons action}\label{SecCSA}

The topological nature of most weakly interacting condensed matter systems is reflected in the Berry connections \cite{Qi2008,Qi2011,Xiao2010}. In particular, a 3d gapped superconductor is shown to be described by an effective gravitational topological action \cite{Wang2011,Ryu2012,Furusaki2013}
\begin{equation}\label{grav}
S_\theta=\frac{\theta}{1536\pi^2}\int\mbox{d}^4x\epsilon^{\mu\nu\rho\sigma}R^{\alpha}_{\ \beta\mu\nu}R^{\beta}_{\ \alpha\rho\sigma}\ ,
\end{equation}
where $R^{\alpha}_{\ \beta\mu\nu}$ is the Riemann tensor of the $3+1$d spacetime, and $\theta$ is the topological theta angle determined by the superconductor up to multiples of $2\pi$. For spacetimes which are spin manifolds, the topological action is quantized to $S_\theta=k\theta$ with $k\in\mathbb{Z}$ being the gravitational instanton number \cite{Wang2011}. In the weak field limit, the Einstein gravitation can be reformulated in the gravitoelectromagnetism framework \cite{Forward1961,Clark2000}, and the above action can be rewritten as \cite{Ryu2012,Furusaki2013}
\begin{equation}
S_\theta=\frac{\theta}{2\pi}\frac{\alpha_g}{2\pi}\int\mbox{d}^4x\bm{E}_g \cdot\bm{B}_g\ ,
\end{equation}
where $\alpha_g$ is the effective coupling constant, while $\bm{E}_g$ and $\bm{B}_g$ are the gravitoelectric field and the gravitomagnetic field, respectively. When a 3d gapped superconductor is TRI, $\exp(iS_{\theta})=\exp(-iS_\theta)$ is required and the theta angle $\theta$ must take $0$ or $\pi$. In general, a TRI topological superconductor characterized by topological number $N_{sc}\in\mathbb{Z}$ will have a theta angle $\theta=N_{sc}\pi \mod 2\pi$ \cite{Qi2010t,Wang2011}. In particular, a superconductor in the $^3$He-B phase has $\theta=\pi$.

This form of gravitational topological action is in direct analogy with the topological electromagnetic action for 3d topological insulators and axion insulators \cite{Qi2008,Li2010,Wang2016}. These gravitational fields are closely related to the thermal transport of the superconductor \cite{Luttinger1964}. In particular, a gradient $\nabla T$ of temperature $T$ can be balanced by a gravitoelectric field $\bm{E}_g=-\nabla T/T$ so that the system can be treated as in equilibrium \cite{Luttinger1964}, while the gravitomagnetic field $\bm{B}_g$ characterizes the rotational energy flow and is proportional to the effective angular velocity of the matter in the system \cite{Ryu2012,Furusaki2013}. As a result, the superconductor exhibits a thermal magneto-electric effect which has a coefficient proportional to the $\theta$ angle.
%

In the band theories of gapped systems, the topological theta angle is given by the momentum-space Chern-Simons action of the non-abelian Berry connection as follows \cite{Qi2008,Qi2011}:
\begin{equation}\label{theta}
\theta=\mathcal{A}_{cs}[a']=\frac{1}{4\pi}\int\text{Tr}\left(a'\wedge da'-i\frac{2}{3}a'\wedge a'\wedge a'\right),
\end{equation}
where $a'$ is as defined in Eq. (\ref{aprime}) for superconductors and is defined similarly for occupied bands of insulators, and the integration is in the whole Brillouin zone.
When the gravitational instanton number $k\neq0$, the total action $S_\theta$ will become a Chern-Simons action at level $k$. However, $\mathcal{A}_{cs}[a']$ cannot be regarded as defining a non-abelian Chern-Simons theory. This is because the Berry connection $a'$ has only U($1$)$^N$ instead of U($N$) gauge freedom, and is generically non-flat for $N>1$: Namely, the non-Abelian Berry curvature $f'=da'-ia'\wedge a'\neq0$ for generic band structures, which is contrary to the classical equation of motion $f'=0$ of the Chern-Simons theory.

To obtain $\theta$ for gapped TRI superconductors from a well-defined momentum-space Chern-Simons theory, we consider the modified Berry connection $a$, and define
\begin{equation}\label{CSa}
\theta=-\mathcal{A}_{cs}[a]=-\frac{1}{4\pi}\int\text{Tr}\left(a\wedge da-i\frac{2}{3}a\wedge a\wedge a\right).
\end{equation}
For 3d gapped TRI superconductors, it can be shown that $\mathcal{A}_{cs}[a]=-\mathcal{A}_{cs}[a']=n\pi$ with $n$ an integer, and the two definitions of $\theta$ coincide (see Appendix \ref{AppBerry}). Formula (\ref{CSa}) has the advantage that $a$ is a well-defined gauge field with U($N$) gauge freedom as we have shown.

More importantly, the equation of motion of Chern-Simons theory, i.e., the flatness of $a$, is satisfied by a natural physical condition which is generically true in mean-field theories of TRI superconductors: two electrons will form a Cooper pair only when they are time-reversal partners and have the same kinetic energy. This condition is equivalent to stating that the two Hermitian matrices $h(\mathbf{k})$ and $\Delta(\mathbf{k})\mathcal{T}$ commute with each other. As a result, they can be simultaneously diagonalized via a unitary transformation as $U_\mathbf{k}h(\mathbf{k})U_\mathbf{k}^\dag=\text{diag}(\epsilon_1(\mathbf{k}),\cdots, \epsilon_N(\mathbf{k}))$ and $U_\mathbf{k}\Delta(\mathbf{k})\mathcal{T}U_\mathbf{k}^\dag=\text{diag}(\Delta_1(\mathbf{k}),\cdots, \Delta_N(\mathbf{k}))$. Therefore, $U_\mathbf{k}$ and $V_\mathbf{k}$ in the SVD of $J(\mathbf{k})$ are related by $V_\mathbf{k}=P_\mathbf{k}U_\mathbf{k}$, where $P_\mathbf{k}=\text{diag}\left(e^{i\phi_1(\mathbf{k})},\cdots,e^{i\phi_N(\mathbf{k})}\right)$ is a diagonal unitary matrix, and $\phi_\alpha(\mathbf{k})$ is the complex phase of $\epsilon_\alpha(\mathbf{k})-i\Delta_\alpha(\mathbf{k})$ $(1\le \alpha\le N)$. Following Eq. (\ref{a}), one can readily show the Berry connection is in the following form:
\begin{equation}\label{flata}
a=iU^\dag\frac{P^\dag dP}{2}U+iU^\dag dU\ ,
\end{equation}
which is just a gauge transformation of $a^U=iP^\dag dP/2=-\text{diag}(d\phi_1,\cdots,d\phi_N)/2$. It is easy to see that $a^U$ as a diagonal matrix of exact $1$-forms is flat, namely, $f^U=da^U-ia^U\wedge a^U=0$. Therefore, we reach the conclusion that $a$ is also flat, namely, the non-abelian field strength (modified Berry curvature) $f=da-ia\wedge a=U^\dag f^U U=0$ at the mean field level.

Unlike insulators where the Berry connection $a'$ is fixed, superconductors always have quantum fluctuations of the pairing amplitude $\Delta(\mathbf{k})$ around the mean-field solution, which lead to fluctuations in both $a'$ and $a$. The pairing fluctuations $\delta\Delta(\mathbf{k})$ generically need not respect the time-reversal symmetry or preserve the commutation of $h(\mathbf{k})$ and $\Delta(\mathbf{k})\mathcal{T}$, therefore leading to non-flat fluctuations in the Berry connection $a$. To the lowest order, the fluctuation of $a$ is linear to $\delta\Delta(\mathbf{k})$.
Explicitly, for the simplest $N=1$ case where $a$ is a U(1) gauge field in 3d, the fluctuation of the phase of $\Delta(\mathbf{k})$ breaks the time-reversal symmetry and gives rise to a $\tau_3$ term in the $2\times2$ Hamiltonian $\widetilde{H}(\mathbf{k})$, which makes the Berry connection $a$ generically non-flat.

In calculating the partition function of the TRI superconductors, all the quantum fluctuations in $\Delta(\mathbf{k})$ around the mean-field solution should be included in the path integral. Correspondingly, any fluctuations in $a$ induced by the pairing fluctuation should be taken into account in evaluating the $\theta$ angle. These fluctuation in $a$ can then be naturally regarded as Gaussian fluctuations in the vicinity of the flat connection which is a classical solution. Since the partition function of the gravitational topological action $S_\theta$ generically contains contributions from topological sectors of all gravitational instanton numbers $k\in\mathbb{Z}$, the action $S_\theta=k\theta$ in each topological sector should be regarded as a physical action. For this reason, we shall view the topological $\theta$ angle as a functional $\theta[a]$ of the gauge field $a$ in the rest of the paper, i.e., as the free energy of the Chern-Simons theory.  More specifically, we will focus on the case where the gravitational instanton pulse has integer unit $k=1$, which leads to level 1 Chern-Simons theory using Eq. (\ref{grav}), and $\theta$ will be identified with the free energy of this level 1 CS theory.

\section{Incorporation of nodal lines}\label{SecInco}

When a TRI superconductor has nodal lines, the system becomes gapless and dissipative, and the gravitational topological action $S_\theta$ does not seem to have a clear physical meaning. On the other hand, nodal lines look like they can behave as Wilson line observables of the Chern-Simons theory.  But what picks the charges of these Wilson lines?  It turns out that these two problems cancel out, as we will show below:  To make $S_\theta$ well defined, we need to gap the system, which can be done by choosing small time-reversal breaking perturbations.  The choices for these perturbation turn out to pick the charges that the Wilson lines carry.  In other words, we will argue that fixing the charges on the Wilson line, which leads to well-defined amplitudes for the CS theory, {\it is equivalent} to choosing small perturbations which gap the system, leading to well defined $\theta$ angle, which will be identified with the free energy of the CS theory in the presence of Wilson loop observables. In this section, we will focus on TRI superconductors without additional symmetries, such as the noncentrosymmetric superconductors, while TRI nodal-line superconductors with inversion symmetry will be discussed in the next section.

\subsection{Nodal lines as Wilson loops}\label{secU1A}

In a TRI superconductor with no other symmetry, the BdG band structure is in general nondegenerate except for some zero-measure submanifolds which generically consists of nodal lines. Therefore, each nodal line is associated with a definite pair of bands which are related by the chiral transformation $\chi$.

It is helpful to consider the minimal toy model with $N=1$ before dealing with the general case, though it is not quite realistic as the spin degrees of freedom (or equivalently, $\mathbf{T}^2=-1$) requires the total number of bands $N$ to be even. The BdG Hamiltonian is then a $2\times2$ matrix
\begin{equation}\label{HU1}
\widetilde{H}(\mathbf{k})=\left(\begin{array}{cc}0&\lambda(\mathbf{k})\\ \lambda^*(\mathbf{k})&0\end{array}\right)\ ,
\end{equation}
where the matrix $J(\mathbf{k})$ from Eq. (\ref{Jk}) reduces to a complex number $\lambda(\mathbf{k})=|\lambda(\mathbf{k})|e^{i\phi(\mathbf{k})}$, with $\phi(\mathbf{k})$ being its complex phase. Accordingly, in the SVD $\lambda(\mathbf{k})= |\lambda(\mathbf{k})|U_\mathbf{k}^\dag V_\mathbf{k}$, one can choose $U_\mathbf{k}=e^{i\gamma(\mathbf{k})}$ and $V_\mathbf{k}=e^{i\phi(\mathbf{k})+i\gamma(\mathbf{k})}$ where $\gamma(\mathbf{k})$ is an arbitrary real function corresponding to the U($1$) gauge freedom, and the Berry connection $a$ reduces to a U($1$) abelian gauge field
\begin{equation}
a(\mathbf{k})=-a'(\mathbf{k})=-\frac{d\phi(\mathbf{k})}{2}-d\gamma(\mathbf{k})\ .
\end{equation}
The nodal lines are characterized by the equation $\lambda(\mathbf{k})=0$, which are closed loops in the momentum space. Therefore, the Berry connection $a$ becomes ill-defined on each nodal line $\mathcal{L}_b$. As one winds around a nodal line $\mathcal{L}_b$ once, the phase $\phi(\mathbf{k})$ will change by $2\pi$ as long as $\lambda(\mathbf{k})$ is a generic continuous function. Besides, the gauge function $\gamma(\mathbf{k})$ must be chosen to change by $2n\pi$ ($n\in\mathbb{Z}$) per winding around the nodal line $\mathcal{L}_b$, so that $U_\mathbf{k}$ and $V_\mathbf{k}$ are single-valued. The gauge field $a$ then satisfies
\begin{equation}\label{flux}
\oint_{\mathcal{C}_b}a=\oint_{\mathcal{C}_b}\bm{a}(\mathbf{k})\cdot\mbox{d}\mathbf{k}=\pi+2n\pi\ ,
\end{equation}
where $\mathcal{C}_b$ is a loop that winds around the nodal line $\mathcal{L}_b$ once, and $n$ is an arbitrary integer. Therefore, a nodal line can be viewed as a vortex line in the momentum space that carries a $\mathbb{Z}_2$ Berry flux $\pi\mod 2\pi$.

Such a U($1$) Berry connection can be derived from the equation of motion of the Chern-Simons theory by regarding the nodal lines as Wilson loops. By modifying the topological theta angle to
\begin{equation}\label{U1theta}
\theta[a]=-\frac{1}{4\pi}\int a\wedge da+\sum_bq_b\oint_{\mathcal{L}_b}a
\end{equation}
and regarding it as a physical action, one finds the equation of motion
\begin{equation}\label{eom}
f=da=\sum_b2\pi q_b\delta^2(\mathbf{k}_\perp-\mathbf{k}_{b})dk_\perp^{1}\wedge dk_\perp^{2}\ ,
\end{equation}
where $\delta^2(\mathbf{k}_\perp-\mathbf{k}_{b})$ stands for the delta function at the nodal line $\mathcal{L}_b$ in the two perpendicular directions $k_\perp^{1}$ and $k_\perp^{2}$ of $\mathcal{L}_b$. The flux formula in Eq. (\ref{flux}) is then satisfied by assigning each Wilson loop a half-integral charge $q_b$.

To better understand such a modified theta angle, we can consider a weak time reversal symmetry breaking in the system. More explicitly, we introduce a small TRI breaking mass term $m$ to the system, so that the Hamiltonian becomes
\begin{equation}
\widetilde{H}'(\mathbf{k})=\left(\begin{array}{cc}m&\lambda(\mathbf{k})\\ \lambda^*(\mathbf{k})&-m\end{array}\right)\ .
\end{equation}
and a nodal line will be lifted into a gap of size $2m$ ($m$ is real). This mass term may emerge effectively from magnetic disorders, or could be induced via uniform tuning of the parameters of the system (e.g., the phase of $\Delta(\mathbf{k})$). If $m$ is nonzero on all nodal lines \cite{mass}, the superconductor becomes fully gapped, and the Berry connections $a$ and $a'=-a$ become well-defined everywhere. In particular, in the limit $m\rightarrow0$, the Berry phase along a loop $\mathcal{C}_b$ that winds around a nodal line $\mathcal{L}_b$ once becomes definite:
\begin{equation}\label{mono}
\oint_{\mathcal{C}_b}a=-\oint_{\mathcal{C}_b}a'=\pi\mbox{sgn}(m)\ ,
\end{equation}
where $\mbox{sgn}(m)$ denotes the sign of $m$. This is analogous to the Berry phase around a Dirac point in the 2d graphene, which falls into $\pm\pi$ depending on the sign of the Dirac mass added. Accordingly, the Wilson loop charge becomes $q_b=\mbox{sgn}(m)/2$ and is no longer indefinite, either. In principle, different nodal lines can have different signs of mass, and therefore different Wilson loop charges. Therefore, the ambiguity of Wilson loop charges $q_b$ disappears once the time reversal symmetry of the superconductor is explicitly weakly broken, and are restricted to $\pm1/2$. We note that $m$ need not be constant, but can be any real function that is nonzero on nodal lines.

Let us clarify the meaning of half-integral $U(1)$ charge.  Recall that we are dealing with a $U(1)$ Chern-Simons theory at level $1$.
The Hilbert space of this theory consists of conformal blocks of the 2d CFT associated with a complex free fermion.\cite{Witten1989}
This theory also admits an extra field, the spin field, which creates branch cuts for the fermion correlations.  Using bosonization, we can view the fermion as $\psi =\exp(i \phi)$ and the spin field as $\sigma^{\pm}=\exp(\pm i \phi/2)$. The insertion of the $\sigma^{\pm}$ fields corresponds to charge $\pm 1/2$ states in the $U(1)$ Chern-Simons theory at level 1.
 Here we have restricted to gravitational instanton number $k=1$.  For a more general $k$, the action is $S_\theta=k\theta[a]$, which we argue emerges from CS theory at level $k$ coupled with Wilson loops: for Eq. (\ref{mono}) to still hold, we also need to rescale the charges on the Wilson loops so that the charge carried by the Wilson loops are now $\pm k/2$.

The gravitational topological action $S_\theta$ with coefficient $\theta[a]$ defined in Eq. (\ref{U1theta}) also acquires a physical meaning under such a weak time reversal symmetry breaking. Since the mass term $m$ gaps out the superconductor, $S_\theta$ is now well-defined as discussed in Sec. \ref{SecCSA}, and one has to go back to Eq. (\ref{theta}) for the definition of $\theta$ in the absence of time reversal symmetry.
On the other hand, making use of the equation of motion in Eq. (\ref{eom}), one can show
\begin{equation}
\sum_bq_b\oint_{\mathcal{L}_b}a=\frac{1}{2\pi}\int a\wedge da\ ,
\end{equation}
and thus the classical value of $\theta[a]$ defined in Eq. (\ref{U1theta}) is
\begin{equation}
\theta_{cl}[a]=\frac{1}{4\pi}\int a\wedge da=\frac{1}{4\pi}\int a'\wedge da'=\mathcal{A}_{cs}(a')\ ,
\end{equation}
which agrees exactly with the definition of $\theta$ in Eq. (\ref{theta}), and we have used the fact that $a=-a'$ for $N=1$. This verifies the validity of the topological theta angle $\theta[a]$ defined in the presence of nodal lines.

As a topological quantum field theory, the Chern-Simons theory has an intrinsic connection with the knot invariants as first revealed by Witten \cite{Witten1989}, which plays an important role in the description of $(2+1)$d topological many-body states such as the quantum Hall states, and quasiparticle statistics therein \cite{Zhang1989,Blok1990,Zhang1992,Moore1991,Lu2012}. Indeed, we will now argue that the partition function of the theory as measured by the $\theta$-angle, in the presence of nodal lines, behaves exactly as expected for a quantum Chern-Simons theory with the corresponding Wilson loops. In the U($1$) case at level $1$, the partition function of CS theory with Wilson loops is given by \cite{Polyakov1988,Witten1989}
\begin{equation}\label{thetalinking}
\theta_{cl}[a]=\pi\sum_{b,c}q_bq_c\Phi(\mathcal{L}_b,\mathcal{L}_c)\ ,
\end{equation}
where
\begin{equation}
\Phi(\mathcal{L}_b,\mathcal{L}_c)=\frac{1}{4\pi}\oint_{\mathcal{L}_b}\mbox{d}x^i\oint_{\mathcal{L}_c}\mbox{d}y^j\epsilon_{ijk}\frac{(x-y)^k}{|x-y|^3}
\end{equation}
is the Gauss linking number between nodal lines $\mathcal{L}_b$ and $\mathcal{L}_c$. In the case of U($1$) Chern-Simons theory, this is also the classical valuation of the action at the critical point (because the action is quadratic). Since the Wilson loop charges $q_b$ are half integral, changing the linking number between two nodal lines by $1$ yields a $\pi/2$ shift in $\theta[a]$. Besides, the calculation of Eq. (\ref{thetalinking}) needs a proper regularization of the self-linking number for $b=c$, which is related to the framing of the Wilson loop.

\begin{figure}[tbp]
  \centering
  \includegraphics[width=3.3in]{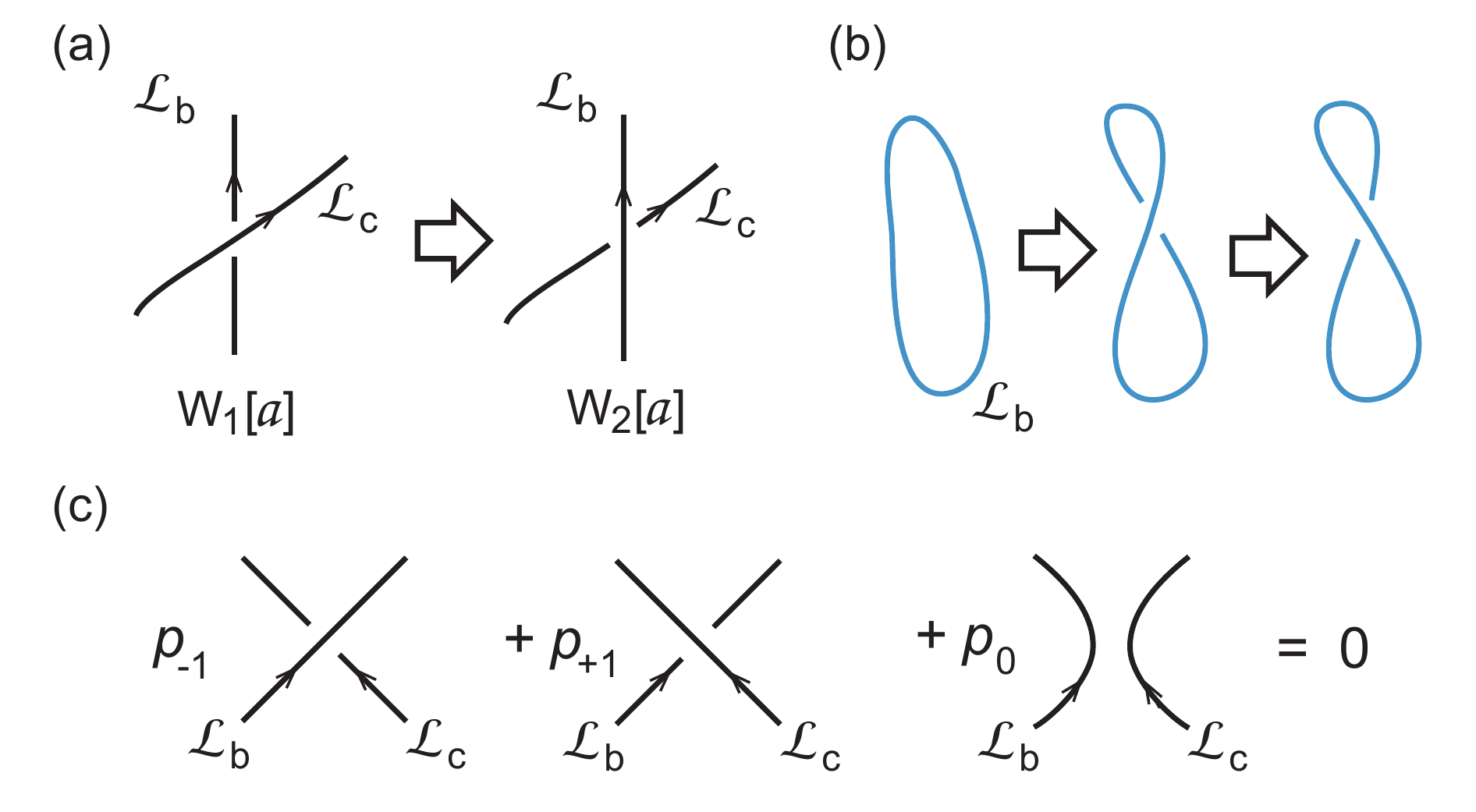}
  \caption{(a) Illustration of local crossing of two nodal lines, which leads to a $\pi/2$ change in the $\theta[a]$ angle. (b) This shows how a nodal line $\mathcal{L}_b$ can cross itself, after which its self-linking number changes by $2$. (c) The generic skein relation for knot polynomials, which gives the rules for the local surgery of a knot configuration.}\label{fig1}
\end{figure}

Now we would like to derive this same result using the definition of the $\theta$-angle for our system. We will show that as we change the linking number between nodal lines, the $\theta$-angle changes exactly as expected for $U(1)$ CS theory at level 1. Consider the local crossing of two nodal lines $\mathcal{L}_b$ and $\mathcal{L}_c$ as shown in Fig. \ref{fig1}(a). $\mathcal{L}_b$ and $\mathcal{L}_c$ can be either different nodal lines or different parts of the same nodal line. The two configurations in Fig. \ref{fig1}(a) can be adiabatically connected by a parameter $t\in[-1,1]$, where $t=-1$ and $t=1$ corresponds to the left and right configurations, respectively. For example, this can be realized by assuming the off-diagonal element in Eq. (\ref{HU1}) to be $\lambda(\mathbf{k})=\lambda_0(k_x+t+ik_y)(k_x+ik_z)$ locally, where $\mathcal{L}_b$ and $\mathcal{L}_c$ are at $(k_x,k_y)=(-t,0)$ and $(k_x,k_z)=(0,0)$, respectively. Accordingly, the field strength in the 4d parameter space $(t,k_x,k_y,k_z)$ is $f=2\pi[q_b\delta(k_x+t)\delta(k_y)d(k_x+t)\wedge dk_y-q_c\delta(k_x)\delta(k_z)dk_x\wedge dk_z]$. Besides, the Wilson loop integrals $\int_{\mathcal{L}_b}a$ and $\int_{\mathcal{L}_c}a$ change by $2\pi q_c$ and $2\pi q_b$, respectively. The resulting total variation in the expectation value of $\theta[a]$ is thus
\begin{equation}\label{calcrossing}
\begin{split}
&\quad\theta_{cl}[a]\Big|_{t=-1}^{t=1}=-\int_{-1}^{1}d\mathcal{A}_{cs}[a]+2\pi q_bq_c+2\pi q_bq_c\\
&=-\frac{1}{4\pi}\int_{4d}\text{Tr}(f\wedge f)+4\pi q_bq_c=-2\pi q_bq_c+4\pi q_bq_c\\
&=2\pi q_bq_c=\pm\frac{\pi}{2}\ \mod 2\pi\ ,
\end{split}
\end{equation}
where we have used the fact that $q_b,q_c$ are $\pm 1/2$. If we define $W[a]=e^{i\theta_{cl}[a]}=\langle e^{i\theta[a]}\rangle$ as a knot invariant, the above result yields a simple skein relation
\begin{equation}
W_1[a]-e^{i2\pi q_bq_c}W_2[a]=0\ ,
\end{equation}
where $W_1[a]$ and $W_2[a]$ correspond to two configurations which are only different locally as shown in Fig. \ref{fig1}(a). This is a special case of the general skein relation as shown in Fig. \ref{fig1}(c), with coefficients $p_{-1}=1$, $p_{+1}=-e^{i2\pi q_bq_c}$ and $p_{0}=0$ \cite{Witten1989,skein}. It is convenient to assign each nodal line $\mathcal{L}_b$ a direction along which the Berry flux is $+\pi$, i.e., $q_b=+1/2$, so that skein relation is unambiguous with $p_{+1}=-i$. In particular, when a nodal line $\mathcal{L}_b$ crosses
a distinct nodal line $\mathcal{L}_c$ the $\theta$-angle changes the contribution to $\theta$-angle changes by $\pi/2$ exactly as is expected from Eq. (\ref{thetalinking}).  This is easy to explain:  both computations can be reduced to finding classical solutions of the CS theory with a given nodal line and evaluating the action at the classical value.  However, we also need to discuss the nodal line crossing itself as shown in Fig. \ref{fig1}(b).  Here $\theta_{cl}[a]$ should also change by $\pi/2$ due to the skein relation, which indicates the `self-linking number' $\Phi(\mathcal{L}_b,\mathcal{L}_b)$ should change by $2$.  The self-linking number is also known as the framing of the knot \cite{Witten1989}, which we have not yet discussed in the context of the nodal lines.  As we will argue below, due to the doubling of the nodal lines due to time reversal invariance, such self-linking, or framing contributions, cancel pairwise as a nodal line crosses itself.  In other words, we can delete the $b=c$ terms in the sum in Eq. (\ref{thetalinking}) as far as its contribution to the total $\theta$-angle is concerned. \cite{framing}

In practice, in TRI superconductors, nodal lines usually occur in pairs at momenta related by the time-reversal symmetry. Therefore, the nodal line crossings are always doubled, and the change in $\theta[a]$ will be doubled to $0$ or $\pi$, depending on the signs of masses of each nodal line in the weak breaking of time-reversal symmetry. For instance, if one assign $\mathcal{L}_b$ and its time-reversal partner the same mass, but assigns $\mathcal{L}_c$ and its time-reversal partner opposite masses, the crossing of $\mathcal{L}_b$ and $\mathcal{L}_c$ (and their time-reversal) will induce a $\pi$ shift in $\theta[a]$.  However, if a nodal line $\mathcal{L}_b$ crosses itself once and recovers the original configuration, due to the time-reversal doubling, time-reversal related nodal lines always have opposite self-crossing phases, and the total change in $\theta_{cl}[a]$ is always zero.  This justifies why we may neglect the contribution of self-linking (or framing number) in finding the contribution of linked nodal lines to $\theta$.

\subsection{Wilson loop in the multi-band case}

Now we proceed to consider nodal lines in general $N$-band TRI superconductors. Following the discussion above Eq. (\ref{flata}), we shall assume $h(\mathbf{k})$ and $\Delta(\mathbf{k})\mathcal{T}$ are simultaneously diagonalized by $U_{\mathbf{k}}=(u_1(\mathbf{k}),\cdots,u_N(\mathbf{k}))^\dag$, where $u_\alpha(\mathbf{k})$ is the $\alpha$-th normalized eigenvector of $h(\mathbf{k})$ ($1\le\alpha\le N$), or the wave function of the $\alpha$-th electron band before superconducting. In the absence of additional symmetries, each nodal line $\mathcal{L}_b$ is associated with a definite electron band. The modified topological theta angle can then be naturally generalized as
\begin{equation}\label{UNtheta}
\theta[a]=-\mathcal{A}_{cs}[a]+\sum_bq_b\oint_{\mathcal{L}_b}u_{\alpha_b}^\dag au_{\alpha_b}\ ,
\end{equation}
where $a$ is now a U($N$) gauge field, $\mathcal{A}_{cs}[a]$ is the non-abelian Chern-Simons action, and $u_{\alpha_b}$ is the $N$-component wave function of the electron band $\alpha_b$ associated with nodal line (or Wilson loop) $\mathcal{L}_b$. Similarly, the Wilson loop charges $q_b$ are half-integral. The Wilson loops $\mathcal{L}_b$ are coupled to the projected U($1$) Berry connection $a^{(\alpha_b)}=u_{\alpha_b}^\dag au_{\alpha_b}$, and thus breaks the U($N$) gauge symmetry. If the 3d momentum space is viewed as a $2+1$d "spacetime", the Wilson loops can be interpreted as world lines of particles, while $u_{\alpha_b}$ are the U($N$) color states of the particles. Due to such a coupling, the Chern-Simons theory governed by $\theta[a]$ acquires more structures. The equation of motion for $a$ then becomes
\begin{equation}\label{eomUN}
f=\sum_b2\pi q_bu_{\alpha_b}u_{\alpha_b}^\dag\delta^2(\mathbf{k}_\perp-\mathbf{k}_{b})dk_\perp^{1}\wedge dk_\perp^{2}\ ,
\end{equation}
where $f$ is now the U($N$) Berry curvature of $a$. When two nodal lines $\mathcal{L}_b$ and $\mathcal{L}_c$ cross each other as shown in Fig. \ref{fig1}(a), a calculation similar to Eq. (\ref{calcrossing}) yields the jump in theta angle
\begin{equation}\label{Njump}
\Delta\theta_{cl}[a]=2\pi q_bq_c|u_{\alpha_b}^\dag u_{\alpha_c}|^2=2\pi q_bq_c\delta_{\alpha_b\alpha_c}\ ,
\end{equation}
provided the eigenvectors $u_{\alpha_b}$ and $u_{\alpha_c}$ are nonsingular (up to phase factors) during the crossing. Therefore, the crossing produces a change in $\theta_{cl}[a]$ only if the two nodal lines are in the same band. We will come back to this point at the end of this subsection.

In metals which become superconductors at low temperatures, nodal lines always live on the 2d fermi surfaces of the metals. One may therefore wonder what kind of fermi surface can give rise to linked nodal lines. Fig. \ref{fig1}(a) shows how two nodal lines of linking number $1$ can be drawn on a $2$-torus fermi surface. We note that they are not time-reversal invariant by themselves. Therefore, their time-reversal partners should coexist on another $2$-torus fermi surface in the BZ. To unlink the two nodal lines $\mathcal{L}_1$ and $\mathcal{L}_2$, one must imagine a process during which the poloidal radius of a certain part of the torus shrinks to zero and then expands back. The two nodal lines can then cross each other at the point of zero poloidal radius.

\begin{figure}[tbp]
  \centering
  \includegraphics[width=3.3in]{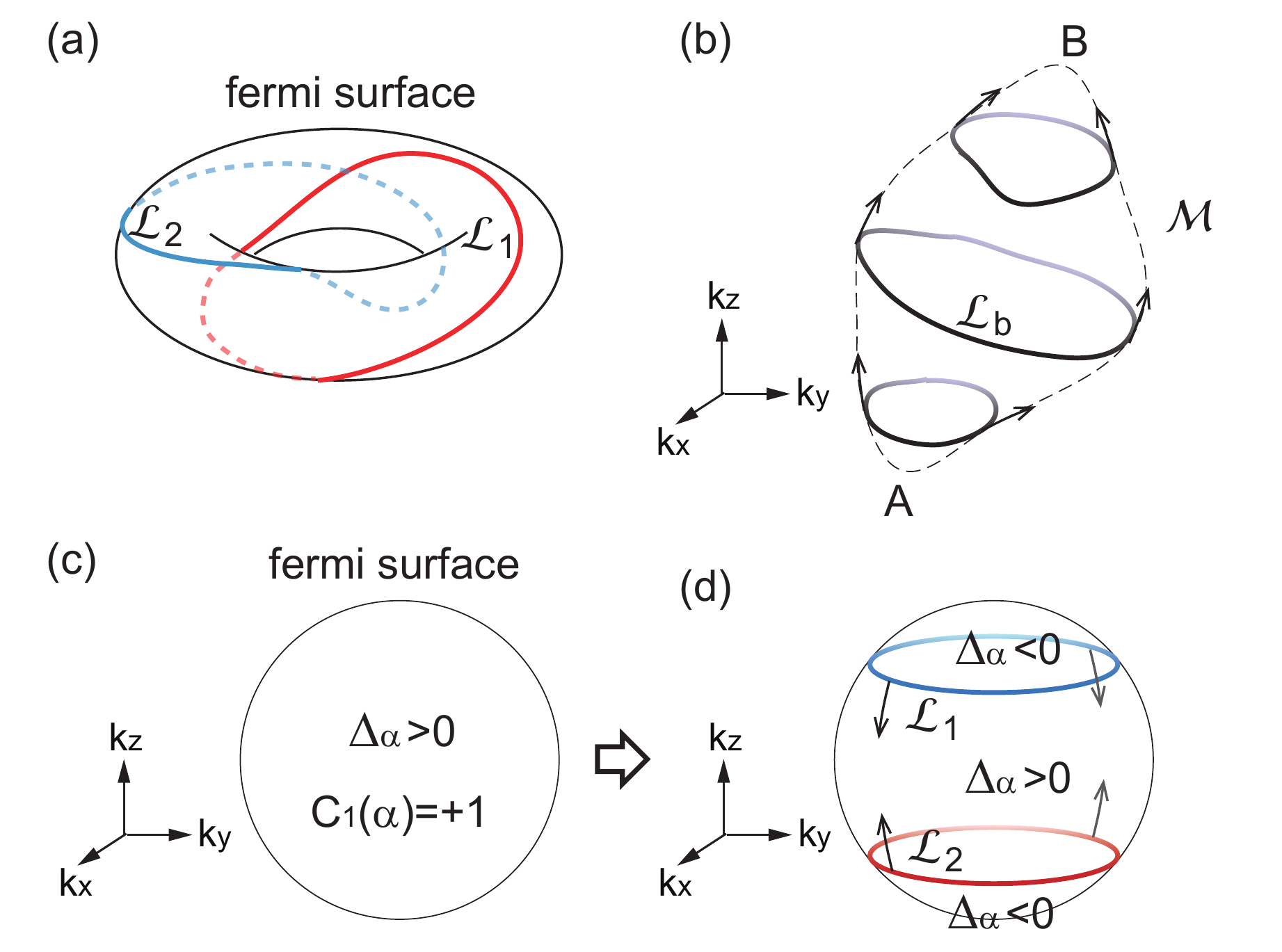}
  \caption{(a) Two linked nodal lines in the same band can be realized on a torus fermi surface. (b) A nodal line can be created at point $A$ and annihilated at point $B$, whose trajectory is a closed surface $\mathcal{M}$. (c)-(d) The sign of pairing $\Delta_\alpha$ on a fermi surface of a gapped TRI superconductor can be reversed by creating nodal lines and making them sweep around the fermi surface, which leads to a phase transition between topological superconductors if the fermi surface Chern number $C_1(\alpha)$ is nonzero.}\label{fig2}
\end{figure}

The theta angle $\theta[a]$ defined in Eq. (\ref{UNtheta}) also explains in a different way the fact that $\theta=N_{sc}\pi \mod 2\pi$ for a TRI gapped topological superconductor with topological number $N_{sc}$. This fact comes from the following observation: consider the process in which a nodal line $\mathcal{L}_b$ is created at point $A$ and then annihilated at point $B$ as shown in Fig. \ref{fig2}(b) as a function of adiabatic parameter $t\in[-1,1]$, during which the system remains TRI. The trajectory of $\mathcal{L}_b$ is thus a closed 2d manifold $\mathcal{M}$ in the 3d momentum space. Accordingly, the change in $\theta_{cl}[a]$ is
\begin{equation}\label{C1}
\begin{split}
&\quad\theta_{cl}[a]\Big|_{t=-1}^{t=1}=-\int_{-1}^{1}d\mathcal{A}_{cs}[a]+q_b\int_{-1}^1d\oint_{\mathcal{L}_b}a^{(\alpha_b)}\\
&=-\frac{1}{4\pi}\int_{4d}\text{Tr}(f\wedge f)+q_b\oint_\mathcal{M}da^{(\alpha_b)}\\
&=0+q_b\oint_\mathcal{M}f^{(\alpha_b)}=2\pi q_b C_1(\alpha_b)\ ,
\end{split}
\end{equation}
where we have used the fact that $f\wedge f=0$ when there is no crossing of nodal lines, while $f^{(\alpha_b)}=da^{(\alpha_b)}=d(u_{\alpha_b}^\dag au_{\alpha_b})$ is the projected U($1$) Berry curvature of electron band $\alpha_b$, and $C_1(\alpha_b)$ is the first Chern number of band $\alpha_b$ on the 2d manifold $\mathcal{M}$. We note that even if the U($N$) Berry curvature $f=0$, the projected U($1$) Berry curvature $f^{(\alpha_b)}$ and thus $C_1(\alpha_b)$ can be generically nonzero. Therefore, $\theta[a]$ of a TRI gapped superconductor can be changed by creation and annihilation of nodal lines.

Physically, this can be done by adiabatically changing the pairing amplitude $\Delta(\mathbf{k})$ of the BdG Hamiltonian. We shall still assume $U_\mathbf{k}h(\mathbf{k})U_\mathbf{k}^\dag=\text{diag}(\epsilon_1(\mathbf{k}),\cdots, \epsilon_N(\mathbf{k}))$ and $U_\mathbf{k}\Delta(\mathbf{k})\mathcal{T}U_\mathbf{k}^\dag=\text{diag}(\Delta_1(\mathbf{k}),\cdots, \Delta_N(\mathbf{k}))$, so a nodal line in band $\alpha$ is given by $\epsilon_\alpha(\mathbf{k})=\Delta_\alpha(\mathbf{k})=0$. We note that $\epsilon_\alpha(\mathbf{k})=0$ defines the fermi surface $\mathcal{M}_\alpha$ of electron band $\alpha$, while $\Delta_\alpha(\mathbf{k})$ is a real function of $\mathbf{k}$. When the superconductor is gapped, $\Delta_\alpha(\mathbf{k})$ is nonzero everywhere on the fermi surface $\mathcal{M}_\alpha$, and thus has a definite sign $\text{sgn}(\Delta_\alpha)$ there. Qi, Hughes and Zhang \cite{Qi2010t} has shown that the topological number $N_{sc}$ of a gapped TRI superconductor is given by
\begin{equation}
N_{sc}=\frac{1}{2}\sum_{\alpha=1}^N C_1(\alpha)\text{sgn}(\Delta_\alpha)\ ,
\end{equation}
where $C_1(\alpha)=\oint_{\mathcal{M}_\alpha}f^{(\alpha)}/2\pi$ is the fermi-surface first Chern number of electron band $\alpha$. Now consider a fermi surface $\mathcal{M}_\alpha$ with $C_1(\alpha)=+1$ and $\Delta_\alpha>0$ as shown in Fig. \ref{fig2}(c) at the beginning. One can then create two nodal lines $\mathcal{L}_1$ and $\mathcal{L}_2$ (which are time-reversal partners) from the north and south poles of the fermi surface as shown in Fig. \ref{fig2}(d), and make them annihilate on the equator. Since $\Delta_\alpha$ on $\mathcal{M}_\alpha$ changes sign across a nodal line, the sign $\mbox{sgn}(\Delta_\alpha)$ is reversed at the end of the process, and $N_{sc}$ decreases by $1$. On the other hand, Eq. (\ref{C1}) tells us that the change in the theta angle $\theta$ is $\Delta\theta=2\pi q_b=-\pi \mod 2\pi$, which therefore agrees with the fact $\theta=N_{sc}\pi$ for TRI topological superconductors. This also verifies that the Wilson loop charges $q_b$ have to be half-integral.

This also indicates that $\theta[a]$ for $N>1$ could acquire a geometric contribution
\[q_b\int d\oint_{\mathcal{L}_b}a^{(\alpha_b)}=q_b\int da^{(\alpha_b)}\]
solely from the deformation of a nodal line $\mathcal{L}_b$, without involving any nodal-line crossings. This is different from the U($1$) Chern-Simons theory discussed in Sec. \ref{secU1A} which is purely topological. Such an additional phase change is due to the Berry curvature carried by the eigenvectors $u_\alpha$. In practice, $\mathbf{T}^2=-1$ always requires the total number of bands $N$ to be even, so such a geometric contribution is in general present and nonzero.

We now go back to Eq. (\ref{Njump}) and add a few more words about it. When two nodal lines $\mathcal{L}_b$ and $\mathcal{L}_c$ cross each other, the kinetic energy matrix $h(\mathbf{k})$ necessarily has two degenerate eigenvectors $u_1$ and $u_2$ of zero eigenvalue at the crossing point, and $u_{\alpha_b}$ and $u_{\alpha_c}$ must approach certain superpositions of $u_1$ and $u_2$ near the crossing point. If the superposition coefficients involve a discontinuous jump before and after the crossing, $u_{\alpha_b}$ and $u_{\alpha_c}$ will be singular, and the jump in theta angle given by Eq. (\ref{Njump}) becomes unclear. However, we can always smoothen the crossing process, i.e., smoothen the discontinuous jump of the superposition coefficients. The change in theta angle is then simply the jump given in Eq. (\ref{Njump}) plus the geometric contribution from the eigenvectors $u_{\alpha_b}$ and $u_{\alpha_c}$ which vary smoothly.

\section{Non-Abelian Nodal Lines}\label{SecU2}

So far, we have discussed TRI superconductors with nondegenerate bands, where nodal lines behave as Wilson loops coupled to projected U($1$) Berry connections. To construct Wilson loops with a non-abelian coupling to the Berry connections, one needs the band structure of the superconductor to have an $N$-fold degeneracy \cite{Wilczek1984}.  Theoretically, we can discuss all $N>1$ cases, but it turns out that the $N=2$ case is particularly well motivated in certain superconductors:  In particular, adding inversion symmetry to the superconductor yields a $2$-fold degeneracy and leads to U($2$) nodal lines, as we shall show below.  In subsection A, we discuss the $U(2)$ case and in the subsection B, we turn to the more general $U(N)$ case.


\subsection{U(2) nodal lines for centrosymmetric superconductors with even-parity pairing}

Many TRI superconductors are also centrosymmetric, namely, have an inversion symmetry $\mathbf{P}$. In this case, the combined symmetry $\mathbf{TP}$ is an anti-unitary symmetry with $(\mathbf{TP})^2=-1$. Since the $\mathbf{TP}$ symmetry keeps the momentum $\mathbf{k}$ invariant, such a symmetry ensures the $N$ electron bands ($N$ is even) to be doubly degenerate according to the Kramers theorem. If one denotes the action of $\mathbf{P}$ on the electron basis as $\mathbf{P}^{-1}\psi_\mathbf{k}\mathbf{P}=\mathcal{P}\psi_{-\mathbf{k}}$ where $\mathcal{P}^2=\mathcal{P}^\dag \mathcal{P}=I_N$, one can show $(\mathcal{TP})^*\mathcal{TP}=-I_N$ and $(\mathcal{TP})^\dag\mathcal{TP}=I_N$, namely, the matrix $\mathcal{TP}$ has the same properties as $\mathcal{T}$. The inversion symmetry of the superconductor requires $\mathcal{P}^\dag h(\mathbf{k})\mathcal{P}=h(-\mathbf{k})$, and $\mathcal{P}^\dag\Delta(\mathbf{k})\mathcal{P}^*=\pm\Delta(-\mathbf{k})$, where $+$ and $-$ signs indicate even and odd parity pairings, respectively. Note that $\Delta(\mathbf{k})=-\Delta^T(-\mathbf{k})$, so this parity requirement can be rewritten as $\Delta(\mathbf{k})\mathcal{P}^*=\mp\mathcal{P}^\dag\Delta^T(\mathbf{k})$, i.e., the matrix $\Delta(\mathbf{k})\mathcal{P}^*$ is anti-symmetric for even parity, and symmetric for odd parity.

When the superconductor has a pairing amplitude $\Delta(\mathbf{k})$ of even parity, nodal lines are allowed to arise and are stable against perturbations. This can be seen as follows. Assume $h(\mathbf{k})=U^\dag_\mathbf{k} E_\mathbf{k}U_\mathbf{k}$, where $E_\mathbf{k}=I_2\otimes \epsilon_\mathbf{k}$ with $\epsilon_\mathbf{k}=\mbox{diag}(\epsilon_{1}(\mathbf{k}),\cdots, \epsilon_{N/2}(\mathbf{k}))$ being a diagonal matrix, while the basis of each $I_2$ ($2\times2$ identity matrix) consists of two eigenstates related by the $\mathbf{TP}$ transformation. Effectively, one can regard the two states as carrying spin up and down, respectively. The matrix $\mathcal{TP}$ can be written as $\mathcal{TP}=U^\dag_\mathbf{k}(i\sigma_2\otimes I_{N/2})U_\mathbf{k}^*$, where $\sigma_{1,2,3}$ denote the Pauli matrices in the doublet basis of $\mathbf{TP}$ related eigenstates. We note that $U_\mathbf{k}$ is not unique. Since $h(\mathbf{k})$ commute with the Hermitian matrix $\Delta(\mathbf{k})\mathcal{T}$, the same matrix $U_\mathbf{k}$ diagonalizes $\Delta(\mathbf{k})\mathcal{T}$ into $2\times 2$ blocks. Together with the condition that $\Delta(\mathbf{k})\mathcal{P}^*=-\Delta(\mathbf{k})\mathcal{T}(\mathcal{TP})^*$ is anti-symmetric (symmetric) for even (odd) parity, we find
\begin{equation}
\Delta(\mathbf{k})\mathcal{T}=U_\mathbf{k}^\dag\left(\begin{array}{ccc}f_1(\mathbf{k})I_2&&\\&\ddots&\\&&f_{N/2}(\mathbf{k})I_2\end{array}\right)U_\mathbf{k}
\end{equation}
for even parity pairing, where $f_n(\mathbf{k})$ are real functions even in $\mathbf{k}$, and
\begin{equation}
\Delta(\mathbf{k})\mathcal{T}=U_\mathbf{k}^\dag\left(\begin{array}{ccc}\bm{\xi}_1(\mathbf{k})\cdot\bm{\sigma}&&\\&\ddots& \\&&\bm{\xi}_{N/2}(\mathbf{k})\cdot\bm{\sigma}\end{array}\right)U_\mathbf{k}
\end{equation}
for odd parity pairing, where $\bm{\xi}_n(\mathbf{k})$ are real vector functions odd in $\mathbf{k}$. When the parity is odd, the gapless condition for the BdG Hamiltonian is $\epsilon_{n}(\mathbf{k})=\bm{\xi}_n(\mathbf{k})=0$, which is overdetermined in 3d and generically has no solution. On the contrary, when the parity is even, the BdG band structure becomes gapless $\epsilon_{n}(\mathbf{k})=f_n(\mathbf{k})=0$, which defines the nodal lines in the doubly-degenerate electron band $\epsilon_{n}$. In practice, such nodal lines widely exist in cuprates with $d$-wave pairings \cite{Tsuei2000} and iron-based superconductors with modulated $s_\pm$-wave pairings \cite{Wangf2011}.

Due to the double degeneracy of the BdG band structure, the nodal lines are coupled to the Berry connection as non-Abelian U($2$) Wilson loops. In the $N=2$ case, the Berry connection $a$ is itself a U($2$) gauge field, and the $\theta$-angle is simply given by the conventional U($2$) non-abelian Chern-Simons theory.
These Wilson loops reside in the fundamental representation $\mathcal{R}$ of U($2$) where the generators are given by $T^{0}=I_2/2$ and $T^{i}=\sigma_i/2$ ($i=1,2,3$) \cite{gene}, and accordingly the Berry connection can be decomposed into $a=\sum_aT^a a^a$. The matrix $J(\mathbf{k})$ in this case is simply $J(\mathbf{k})=h(\mathbf{k})-i\Delta(\mathbf{k})\mathcal{T}=\lambda(\mathbf{k})I_2$, where $\lambda(\mathbf{k})$ is a complex number function. A nodal line is defined by $\lambda(\mathbf{k})=0$, while winding around the nodal line increases the phase $\phi(\mathbf{k})$ of $\lambda(\mathbf{k})$ by $2\pi$. To compute the classical value of gauge field $a$, one can choose $U_\mathbf{k}=e^{-i\phi(\mathbf{k})}I_2$ and $V_\mathbf{k}=I_2$ in the vicinity of the nodal line, in which case the field strength is
\begin{equation}\label{f0}
f=2\pi T^{0}\delta^2(\mathbf{k}_\perp-\mathbf{k}_{b})dk_\perp^{1}\wedge dk_\perp^{2}\ .
\end{equation}
Alternatively, one can choose $U_\mathbf{k}=e^{i(-T^0+\hat{n}^iT^i)\phi(\mathbf{k})}$ and $V_\mathbf{k}=e^{i(T^0+\hat{n}^iT^i)\phi(\mathbf{k})}$ with a unit vector $\hat{\bm{n}}=(\hat{n}^1,\hat{n}^2,\hat{n}^3)$, which yields a field strength
\begin{equation}\label{fi}
f=2\pi \hat{n}^iT^{i}\delta^2(\mathbf{k}_\perp-\mathbf{k}_{b})dk_\perp^{1}\wedge dk_\perp^{2}\ .
\end{equation}
Both choices ensure $U_\mathbf{k}$ and $V_\mathbf{k}$ to be single-valued. This choice of classical solutions exactly results from the U($2$) gauge freedom of the nodal line. Similar to the U($1$) case, the field strength of a nodal line becomes well defined when the time reversal symmetry or the inversion symmetry of the system is weakly broken. For instance, the Hamiltonian could acquire a small mass term and become
\begin{equation}
\widetilde{H}'(\mathbf{k})=\left(\begin{array}{cc}m_0I_2+m\hat{\bm{n}}\cdot\bm{\sigma}&\lambda(\mathbf{k})I_2\\ \lambda^*(\mathbf{k})I_2&-m_0I_2-m\hat{\bm{n}}\cdot\bm{\sigma}\end{array}\right)\ .
\end{equation}
Explicitly, there are four degrees of freedom, which are $m_0$ and $m_i \hat n_i$ ($i = 1,2,3$), which is the expected number for arbitrary fluctuations of a U($2$) gauge field in $d = 3$ because $N^2(d - 2) = 4$, where here $N = 2$ and $d = 3$.

It is important to note that the field strength $f$ will depend on the signs of the eigenvalues of the upper $2\times2$ block.  If the signs of the mass eigenvalues of the upper $2\times2$ block is $(\pm,\pm)$, this leads to the curvature sitting in the abelian subalgebra of $U(2)$ given by
\begin{equation}\label{fsigns}
f=2\pi \left[{\pm 1\over 2},{\pm 1\over 2}\right]\  \delta^2(\mathbf{k}_\perp-\mathbf{k}_{b})dk_\perp^{1}\wedge dk_\perp^{2}.
\end{equation}
This can be interpreted (when the gravitational instanton number $k=1$) as the response of the $U(2)$ system to a charged object
with charges $[{\pm 1/2},{\pm 1/2}]$ in the Cartan of $U(2)$.  If both signs are the same, the Wilson loop will carry only a $U(1)\subset U(2)$ charge.  If the signs are opposite, the two possibilities correspond to the fundamental of $SU(2)\subset U(2)$ together with some overall $U(1)$ charge.  Note that when the inversion symmetry is broken, the U($2$) nodal lines naturally breaks down to abelian nodal lines with Wilson loop charges $\pm1/2$, in agreement with our observation in Sec. \ref{SecInco}.  The Wilson loop observables etc. will work exactly as in the abelian case we already discussed, because the level of the $U(2)$ CS theory is 1.

If we change the gravitational instanton number so that the level of the $U(2)$ CS theory is changed to $k$, we would need to choose different weights in the $U(2)$ weight lattice in order to get the same $f$ we computed in Eq. (\ref{fsigns}).  In particular, the Cartan weights will now be
$(\pm k/2, \pm k/2)$.  Apart from the $U(1)$ part, this belongs to the $k$-fold symmetric representation of the fundamental of $SU(k)$, which is known to generate a free current and thus means that its correlations behave as in the abelian case, leading exactly to the same result expected, where $\theta $ gets multiplied by a factor $k$.

\subsection{U($N$) nodal lines}

Though uncommon in nature, condensed systems with $n$-fold degenerate ($n>2$) band structures are theoretically possible when the symmetries of the systems are high enough. For instance, systems where electrons carry both a spin $1/2$ and a pseudospin $1/2$ may have a $4$-fold degenerate band structure when there are no spin-pseudospin couplings. When such a system is TRI and develops a superconductivity, nodal lines with a U($n$) symmetry may arise.

For simplicity, here we only consider the case when the total number of electron bands is $N=n$, which is the minimal system that realizes U($n$) nodal lines. The matrix $J(\mathbf{k})$ in Eq. (\ref{Jk}) is then of the form $J(\mathbf{k})=\lambda(\mathbf{k})I_N$ with $\lambda(\mathbf{k})=|\lambda(\mathbf{k})|e^{i\phi(\mathbf{k})}$ being a complex function. A nodal line is defined by $\lambda(\mathbf{k})=0$; it occupies the fundamental representation of U($N$) group, of which the generators are $T^0=I_N/\sqrt{2N}$ and the $N^2-1$ generators $T^i$ of SU($N$) group \cite{gene}. The matrices $U$ and $V$ around the nodal line can be generically chosen as
\begin{equation}
V_\mathbf{k}=e^{i\phi(\mathbf{k})}U_\mathbf{k}=e^{i\omega^aT^a\phi(\mathbf{k})}\ ,
\end{equation}
where $\omega^0=\sqrt{2/N}\ell$ ($\ell\in\mathbb{Z}$), and $\omega^i$ ($1\le i\le N^2-1$) generates the element $e^{i\omega^iT^i}=e^{-i2\pi\ell/N}$ of the center of SU($N$). The resulting field strength $f$ is
\begin{equation}
f=2\pi \omega^aT^{a}\delta^2(\mathbf{k}_\perp-\mathbf{k}_{b})dk_\perp^{1}\wedge dk_\perp^{2}\ ,
\end{equation}
which is a classical solution of the U($N$) Chern-Simons theory with U($N$) Wilson loops.   In paricular, just as in the $U(2)$ case this will lead to curvature which is proportional to $[{\pm 1\over 2},...,{\pm 1\over 2}]$, which is in the fundamental representation of $U(N)$ at level 1, leading to similar formulae as before.

\section{nodal lines induced by line defects and topological strings}\label{Secdefect}

In many cases it is desirable to consider the effect of defects in the physical space of a condensed matter system. In the physical space of 3d superconductors, vortex lines can be naturally created as line defects, while crystallographic line defects (dislocations and disclinations) may also be present, which lead to a breaking of the translational symmetry. Interestingly, as we shall show below, such line defects in physical space give rise to effective nodal lines in the BZ, or more precisely, in the phase space under a semiclassical approximation.

For simplicity, let us consider a line defect along $z$-direction located at $x=y=0$ in the physical space. The line defect will generically produce a potential energy $V_d(x,y)$ for the BdG quasi-particles that is centered at $x=y=0$. An quasi-particle with kinetic energy $\epsilon(\mathbf{k})$ will then have a Hamiltonian:
\begin{equation}
H_d(k_z)=\epsilon(-i\partial_x,-i\partial_y,k_z)+V_d(x,y) \ ,
\end{equation}
where we have used the fact that $k_z$ is still a good quantum number. Assume the Hamiltonian has a bound state $|\psi_d(k_z)\rangle$ with a bound state energy $\epsilon_d(k_z)$ \cite{Ran2009,Slager2013,Slager2014}. Such a bound state is in general a wave packet of size $\ell_d$ centered at $x=x_0(k_z)$ and $y=y_0(k_z)$ in the physical space, and also a wave packet of size $\pi/\ell_d$ with the center located at $k_x=k_{x0}(k_z)$ and $k_y=k_{y0}(k_z)$ in the BZ, where $\ell_d$ is the localization length of the bound state (which we assume is much larger than the lattice constants), and the functions $x_0(k_z)$, $y_0(k_z)$, $k_{x0}(k_z)$ and $k_{y0}(k_z)$ are determined by the line defect Hamiltonian $H_d(k_z)$. Therefore, all the bound states on the line defect live on a "smeared out" line in the BZ given by $(k_x,k_y)=(k_{x0}(k_z),k_{y0}(k_z))$, which is a closed loop, and also live on a "smeared out" line $(x,y)=(x_0(k_z),y_0(k_z))$ in the Lagrangian space $(x,y,k_z)$ \cite{Lspace}.

Now we assume a quasi-particle is bounded on the line defect. We then introduce an additional slow potential $V_s(z)\ll |V_d(x,y)|$ which slightly breaks the translational symmetry along the line defect. Such a slow potential could result from inhomogeneity of the system or slight distortions of the defect line. Consider a local minimum where $V_s(z)\approx z^2/2M$. The quasi-particle is then located near $z=0$, and has an effective $1$d Hamiltonian
\[H_q=\frac{(id/dk_z+\bm{v}_K\cdot\bm{a}+\bm{v}_L\cdot\widetilde{\bm{a}})^2}{2M}+\epsilon_d(k_z)\ ,\]
where we have defined $\bm{v}_K=(dk_{x0}/dk_z,dk_{y0}/dk_z,1)$ and $\bm{v}_L=(dx_0/dk_z,dy_0/dk_z,1)$, while $\bm{a}$ is the Berry connection in the BZ defined in Sec. II. and we have also included a new Berry gauge field $\widetilde{\bm{a}}(x,y,k_z)=(\widetilde{a}_{x},\widetilde{a}_{y},\widetilde{a}_{k_z})$ in the Lagrangian space $(x,y,k_z)$ \cite{Ooguri2000,Lspace}. Such a Berry gauge field $\widetilde{\bm{a}}$ arises generically when the translational symmetry of the system in the $x$-$y$ plane is broken \cite{Bulmash2015}. For instance, in the presence of $z$-direction translational invariant lattice displacements, a particle may acquire an adiabatic $z$-direction translation $z_t(x,y)$ when moving in the $x$-$y$ plane, thus gain a phase factor $e^{ik_zz_t}$, which induces a nontrivial Berry gauge field $(\widetilde{a}_{x},\widetilde{a}_{y})=(k_z\partial_xz_t,k_z\partial_yz_t)$ with field strengths $\widetilde{f}_{k_zx}=\partial_xz_t$ and $\widetilde{f}_{k_zy}=\partial_yz_t$. Furthermore, if there are also dislocations, $z_t$ will depend the electron's moving path in the $x$-$y$ plane, and a nonzero $\widetilde{f}_{xy}$ proportional to $k_z$ will arise.

Note that the first term in the Hamiltonian $H_q$ can be viewed as a kinetic energy in the $k_z$ space, while the second term $\epsilon_d(k_z)$ behaves as a periodic potential along $k_z$. In this perspective, the Lagrangian corresponding to $H_q$ is
\[L_q(k_z,\dot{k}_z)=\frac{M}{2}\dot{k_z}^2+\dot{\mathbf{K}_q}\cdot(\bm{a}+\widetilde{\bm{a}})-\epsilon_d(k_z)\ ,\]
where $\mathbf{K}_q=(x_0(k_z),y_0(k_z),0,k_{x0}(k_z),k_{y0}(k_z),k_z)$ is the position of the particle in the 6d phase space, and $\dot{\mathbf{K}}_q$ is the time derivative of $\mathbf{K}_q$. Since the $z$ coordinate of the particle is not important, we simply fix it to its mean value $z=0$. All the possible positions $\mathbf{K}_q (k_z)$ forms a loop $\mathcal{L}_d$ in the phase space $(x,y,z,k_x,k_y,k_z)$.
In the periodic time path integral formalism, the quasi-particle can wind around the loop $n_w$ times, of which the tunnelling amplitude is given by $\exp(-|n_w|S_t)\approx\exp(-|n_w|\oint_{\mathcal{L}_d}\sqrt{2M\epsilon_d(k_z)}dk_z)$ according to the WKB approximation. The path integral of the bounded quasi-particle is thus
\begin{equation}\label{sum}
Z_q=\int\mathcal{D}\mathbf{k}(t)e^{i\int L_q dt}\approx\sum_{n_w} e^{in_w\oint_{\mathcal{L}_d} (a+\widetilde{a})-|n_w|S_t}.
\end{equation}
Therefore, one sees the loop $\mathcal{L}_d$ plays exactly the role of a Wilson loop (nodal line) that couples to both $a$ and $\widetilde{a}$, where the Wilson loop charge is now the winding number $n_w$. We note that the shape of the slow potential $V_s(z)$ is only relevant to the tunnelling action $S_t$, while does not affect the Wilson loop integral.
In this case, the Wilson loop charge $n_w$ and the level of CS action of $a$, i.e., the gravitational instanton number $k$, are independent of each other.
In principle, the gauge field $\widetilde{a}$ may also have a physical CS action at some level $\widetilde{k}$ in the Lagrangian space $(k_z,x,y)$, which then yields a more interesting doubled CS theory.

Generically, the number of quasi-particles bounded on the line defect is not limited to one. To take all number of quasi-particles into account, we can second quantize the above action, rewriting the path integral as
\begin{equation}\label{mohem}
Z_q=\int\mathcal{D}\bar{\psi}_{j}\mathcal{D}{\psi}_{j}e^{i\int dt\bar{\psi}_j[i\partial_t+\dot{\mathbf{K}}_q^j(t)\cdot(\bm{a}+\widetilde{\bm{a}})+iS_t/\beta]\psi_j}\ ,
\end{equation}
where $\psi_j$ is the fermion mode associated with a particular closed path $\mathbf{K}_q^j(t)$ on the loop $\mathcal{L}_d$, $t$ is the time, and $\beta$ is the period in the time direction.

We note that the Berry connection $a$ here should be treated as a gauge field defined at the specific point $(x,y,z)=(x_0(k_z),y_0(k_z),0)$ of the physical space where the quasi-particles are trapped. In fact, the strict definition of the Berry connection $a$ does require specifying a point $\mathbf{r}$ in the physical space as an origin \cite{Freed2013}, namely, $a$ is defined via the Bloch wave functions in the unit cell at $\mathbf{r}$. In particular, since the Bloch wave functions have a shift ambiguity $|\alpha,\mathbf{k}\rangle\rightarrow e^{i\mathbf{k}\cdot\mathbf{R}}|\alpha,\mathbf{k}\rangle$ under a translation $\mathbf{r}\rightarrow\mathbf{r}+\mathbf{R}$ (where $\mathbf{R}$ is a lattice vector), the Berry connections $a$ defined at $\mathbf{r}$ and $\mathbf{r}+\mathbf{R}$ are related by $a_{\mathbf{r}+\mathbf{R}}=a|_\mathbf{r}+\mathbf{R}\cdot d\mathbf{k}$. This shift ambiguity is closely related to the 3d quantum Hall effect \cite{Kohmoto1992,Freed2013}, but does not affect the calculation of $\theta[a]$ of TRI systems here. Similarly, the gauge field $\widetilde{a}$ and the associated Lagrangian space $(k_z,x,y)$ should be viewed as located at the specific position $z=0$ (local minimum of $V_s(z)$) and $(k_x,k_y)=(k_{0x}(k_z),k_{0y}(k_z))$ \cite{Bulmash2015,Ran2009}.


It turns out that the above description is very natural in the context of topological strings, which was introduced by Witten \cite{Witten:1992fb} and gives a realization of Chern-Simons theory as a string theory.  This corresponds to strings propagating on a six dimensional symplectic manifold which is typically taken to be a Calabi-Yau 3-fold. $N$ D-branes wrapping Lagrangian 3-cycles lead to $U(N)$ Chern-Simons theory living on it.  Moreover, intersecting D-branes along a loop leads to insertion of Wilson Loop observables as in Eq. (\ref{sum}) on the intersection \cite{Ooguri2000}.   To connect this to the present discussion, we note that the phase space $(x,y,z,k_x,k_y,k_z)$ is a sympletic manifold $T^*T^3$.  We take two Lagrangians to be one wrapped around $(k_x,k_y,k_z)$ giving us the CS theory with gauge field $\bm{a}$, and the other to represent the defect, given by the Lagrangian $(k_z,x,y) $ with gauge field $\widetilde{\bm{a}}$ living on it.
These Lagrangians intersect along $k_z$, leading to a fermion field  which couples to the two gauge fields exactly as in Eq. (\ref{mohem}) \cite{Ooguri2000}.
It is remarkable that phase space in 3d can naturally give a realization of topological strings!

Though our discussion here is based on a straight line defect in physical space, the result holds for generic smooth line defects. In the semiclassical approximation, one can define a momentum along the defect line which plays the role of $k_z$ here, and all the rest of the derivation will follow.

\section{Discussion}

We have seen how a modified Berry's connection of a TRI superconductor can behave as a fluctuational gauge field in the 3d BZ governed by a CS theory whose level is given by the gravitational instanton number in the physical spacetime. Moreover, we have seen that gapless nodal lines play the role of Wilson loop observables for the Chern-Simons theory.
The free energy of this CS theory computes the $\theta$-angle of the topological gravitational response $R\wedge R$ of this system. Changing of the topological mutual linking numbers of nodal lines leads to a shift of the $\theta$-angle in units of $\pi$.
Whereas the linking of degenerate manifolds is also studied for the 5d Weyl semimetals \cite{Lian2016}, it is reasonable to expect linking and knot invariants to play an important role in more generic gapless topological states of matter.

The examples we have discussed exhibit mostly an abelian structure. Even in the multi-band case where we got $U(N)$ CS theory, the computations reduce to $U(1)^N$ abelian CS theory. It would be nice to find ways where the linked nodal amplitudes are genuinely non-abelian.

This paper provides an example of how the Berry connection in the BZ can become fluctuating. In addition, in the discussion of physical space defects in Sec. \ref{Secdefect}, we see possibility of realizing topological string theory in a condensed matter set-up. We have seen an intriguing realization of topological strings on the six dimensional phase space where defects play the role of D-branes. It would be very interesting to develop this connection further, for instance, to see how the gauge field $\widetilde{a}$ in the Lagrangian manifold $(k_z,x,y)$ can also be made dynamical and governed by a CS action. It would also be useful to find more examples along these lines in different condensed matter systems in different dimensions, which need not be limited to topological theories.

\begin{acknowledgments}
We would like to thank C. Liu and C. Xu for discussions. We gratefully acknowledge support from the Simons Center for Geometry and Physics,
Stony Brook University, where some of the research for this paper was performed
during the 2016 Simons Summer Workshop. CV would also like to thank Stanford Institute for Theoretical Physics for hospitality during part of this work. The research of BL and SCZ is supported by the NSF grant DMR-1305677. The research of CV is supported by the NSF grant PHY-1067976. The research of FV is supported by the NSF grant DMR-1151208.
\end{acknowledgments}

\appendix

\section{Relation between two kinds of Berry connections}\label{AppBerry}

The conventionally defined Berry connection $a'_{\alpha\beta}(\mathbf{k})=i\langle\alpha,\mathbf{k}|d|\beta,\mathbf{k}\rangle$ in the literature \cite{Qi2008,Qi2011,Xiao2010} provides a non-abelian generalization of the U($1$) Berry connection proposed by Berry \cite{Berry1984}, where $|\alpha,\mathbf{k}\rangle$ are the eigenstates of the Hamiltonian $H$ and $\alpha, \beta$ runs over the $N$ occupied bands. In this definition, the number of occupied bands and that of empty bands are not required to be equal. When the system is gapped, such a Berry connection naturally describes the topology of the $N$ occupied bands, which remains unchanged as long as the gap of the system does not close. However, one should note that the generic U($N$) gauge transformation of $a'$ necessarily involves the variation of the Hamiltonian, therefore is not a symmetry of the system. This is because a gauge transformation involves a redefinition of eigenstates in the space spanned by the original $N$ occupied eigenstates $|\alpha,\mathbf{k}\rangle$, which varies the Hamiltonian if the $N$ eigenstates are not all degenerate. For example, in the BdG Hamiltonian of Eq. \ref{Hdiag} we considered here, a gauge transformation $a'\rightarrow w^\dag aw+iw^\dag dw$ with an $N\times N$ unitary matrix $w(\mathbf{k})$ is realized by $V\rightarrow w^\dag V$ and $U\rightarrow w^\dag U$, under which the Hamiltonian $\widetilde{H}(\mathbf{k})$ in Eq. (\ref{Hdiag}) becomes
\begin{equation}
\widetilde{H}_w(\mathbf{k})= \Lambda_\mathbf{k}\left(\begin{array}{cc}wD_\mathbf{k}w^\dag&\\&-wD_\mathbf{k}w^\dag\end{array}\right)\Lambda^\dag_\mathbf{k}\ .
\end{equation}
Therefore, the Hamiltonian is varied if $w$ does not commute with $D_\mathbf{k}$, which is not a unitary transformation of the Hamiltonian. When all the bands are nondegenerate, the Hamiltonian is invariant only if $w$ is diagonal, i.e., one is making a U($1$)$^N$ transformation. In contrast, the projector onto the $N$ occupied bands $\hat{\bm{\Pi}}=\sum_{\alpha=1}^N|\alpha,\mathbf{k}\rangle\langle\alpha,\mathbf{k}|$ is invariant under an arbitrary gauge transformation of $a'$, so one may think of the gauge symmetry of $a'$ as a symmetry of the projector.


For the purpose of this paper, we wish to find a Berry connection with a gauge symmetry that is the symmetry of the Hamiltonian. Here we shall show the modified Berry connection $a_{mn}(\mathbf{k})=i\sum_\alpha\langle0|\psi_{m,\mathbf{k}}|\alpha,\mathbf{k}\rangle d\langle \alpha,\mathbf{k}|\psi_{n,\mathbf{k}}^\dag|0\rangle$ is such a well-defined U($N$) gauge field, which is uniquely defined for superconductors (or any Hamiltonian invariant under the chiral transformation). To see this, we note that under a U($N$) unitary transformation of the electron basis $\psi_\mathbf{k}$ used for writing the Hamiltonian and the eigenstates
\[\psi_\mathbf{k}\rightarrow g(\mathbf{k})\psi_\mathbf{k}\]
where $g$ is an $N\times N$ unitary matrix, the hole basis transforms as
\[\mathcal{T}^*\psi_{-\mathbf{k}}^{\dag T}\rightarrow
=-\left(\psi^\dag_{-\mathbf{k}}\mathcal{T}^\dag g^{*\dag}\right)^T=g\mathcal{T}^*\psi_{-\mathbf{k}}^{\dag T}\ .\]
This leads to $V\rightarrow Vg$ and $U\rightarrow Ug$, and transforms $a$ given in Eq. (\ref{a}) as
\begin{equation}
a\rightarrow g^\dag ag+ig^\dag dg\ ,
\end{equation}
which is exactly the gauge transformation of the U($N$) gauge field. Meanwhile, the Hamiltonian is solely doing a unitary transformation. Therefore, the basis transformation $\psi_\mathbf{k}\rightarrow g(\mathbf{k})\psi_\mathbf{k}$ gives the U($N$) gauge transformation of $a$, which is a trivial "symmetry" of the Hamiltonian.

One may wonder what happens to the conventional Berry connection $a'$ if one makes a unitary basis transformation $\psi_\mathbf{k}\rightarrow g(\mathbf{k})\psi_\mathbf{k}$. As we have seen, such a transformation leads to $V\rightarrow Vg$ and $U\rightarrow Ug$, and according to Eq. (\ref{aprime}), $a'$ will transform as
\begin{equation}
a'\rightarrow a'+\frac{i}{2}[U(gdg^\dag)U^\dag+V(gdg^\dag)V^\dag]
\end{equation}
which is not a gauge transformation. Therefore, the basis transformation does not transform $a'$ legally. Thus we conclude the conventional Berry connection $a'$ is only allowed to do U($1$)$^N$ gauge transformation if the Hamiltonian is fixed, while the modified Berry connection $a$ has the full U($N$) gauge freedom.

Now we proceed to show the two Berry connections have opposite Chern-Simons actions $\mathcal{A}[a]=-\mathcal{A}[a']$ for TRI gapped superconductors. From the expressions of Eq. (\ref{aprime}) and Eq. (\ref{a}), we have
\begin{widetext}
\begin{equation}
\begin{split}
&\text{Tr}\left(a\wedge da-i\frac{2}{3}a\wedge a\wedge a\right)=\text{Tr}\Big[-\frac{(U^\dag dU+V^\dag dV)\wedge(dU^\dag\wedge dU+ dV^\dag\wedge dV)}{4}
-\frac{2}{3}\frac{(U^\dag dU+V^\dag dV)^3}{8}\Big]\\
&\qquad=\frac{1}{6}\text{Tr}\big(U^\dag dU\wedge U^\dag dU\wedge U^\dag dU+V^\dag dV\wedge V^\dag dV\wedge V^\dag dV\big)\ ,
\end{split}
\end{equation}
where we have used the fact that $dU^\dag=-U^\dag dU U^\dag$. Following the same calculation one will find
\begin{equation}
\begin{split}
&\text{Tr}\left(a'\wedge da'-i\frac{2}{3}a'\wedge a'\wedge a'\right)=\frac{1}{6}\text{Tr}\big(U dU^\dag\wedge U dU^\dag\wedge U dU^\dag+V dV^\dag\wedge V dV^\dag\wedge V dV^\dag\big)\\
&\qquad=-\frac{1}{6}\text{Tr}\big(U^\dag dU\wedge U^\dag dU\wedge U^\dag dU+V^\dag dV\wedge V^\dag dV\wedge V^\dag dV\big)\ .
\end{split}
\end{equation}
\end{widetext}
Therefore, for gapped superconductors where $U$ and $V$ can be well defined everywhere in the Brillouin zone, we have $\mathcal{A}[a]=-\mathcal{A}[a']$. Besides, note that the integral in the 3d Brillouin zone
\begin{equation}
\frac{1}{24\pi^2}\int \text{Tr}\left( U dU^\dag\wedge U dU^\dag\wedge U dU^\dag\right)
\end{equation}
is an integer for unitary matrix $U$, we conclude that $\mathcal{A}[a]=-\mathcal{A}[a']=n\pi$ with $n\in\mathbb{Z}$, in agreement with the requirement of the time reversal symmetry. In the presence of nodal lines, however, $U$ and $V$ become ill-defined on the nodal lines, and this conclusion no longer holds.



\begin{thebibliography}{54}
\expandafter\ifx\csname natexlab\endcsname\relax\def\natexlab#1{#1}\fi
\expandafter\ifx\csname bibnamefont\endcsname\relax
  \def\bibnamefont#1{#1}\fi
\expandafter\ifx\csname bibfnamefont\endcsname\relax
  \def\bibfnamefont#1{#1}\fi
\expandafter\ifx\csname citenamefont\endcsname\relax
  \def\citenamefont#1{#1}\fi
\expandafter\ifx\csname url\endcsname\relax
  \def\url#1{\texttt{#1}}\fi
\expandafter\ifx\csname urlprefix\endcsname\relax\def\urlprefix{URL }\fi
\providecommand{\bibinfo}[2]{#2}
\providecommand{\eprint}[2][]{\url{#2}}

\bibitem[{\citenamefont{Zhang et~al.}(1989)\citenamefont{Zhang, Hansson, and
  Kivelson}}]{Zhang1989}
\bibinfo{author}{\bibfnamefont{S.~C.} \bibnamefont{Zhang}},
  \bibinfo{author}{\bibfnamefont{T.~H.} \bibnamefont{Hansson}},
  \bibnamefont{and} \bibinfo{author}{\bibfnamefont{S.}~\bibnamefont{Kivelson}},
  \bibinfo{journal}{Phys. Rev. Lett.} \textbf{\bibinfo{volume}{62}},
  \bibinfo{pages}{82} (\bibinfo{year}{1989}).

\bibitem[{\citenamefont{Blok and Wen}(1990)}]{Blok1990}
\bibinfo{author}{\bibfnamefont{B.}~\bibnamefont{Blok}} \bibnamefont{and}
  \bibinfo{author}{\bibfnamefont{X.~G.} \bibnamefont{Wen}},
  \bibinfo{journal}{Phys. Rev. B} \textbf{\bibinfo{volume}{42}},
  \bibinfo{pages}{8145} (\bibinfo{year}{1990}).

\bibitem[{\citenamefont{Zhang}(1992)}]{Zhang1992}
\bibinfo{author}{\bibfnamefont{S.~C.} \bibnamefont{Zhang}},
  \bibinfo{journal}{Int. J. Mod. Phys. B} \textbf{\bibinfo{volume}{06}},
  \bibinfo{pages}{803} (\bibinfo{year}{1992}).

\bibitem[{\citenamefont{Moore and Read}(1991)}]{Moore1991}
\bibinfo{author}{\bibfnamefont{G.}~\bibnamefont{Moore}} \bibnamefont{and}
  \bibinfo{author}{\bibfnamefont{N.}~\bibnamefont{Read}},
  \bibinfo{journal}{Nucl. Phys. B} \textbf{\bibinfo{volume}{360}},
  \bibinfo{pages}{362 } (\bibinfo{year}{1991}), ISSN \bibinfo{issn}{0550-3213}.

\bibitem[{\citenamefont{Lu and Vishwanath}(2012)}]{Lu2012}
\bibinfo{author}{\bibfnamefont{Y.-M.} \bibnamefont{Lu}} \bibnamefont{and}
  \bibinfo{author}{\bibfnamefont{A.}~\bibnamefont{Vishwanath}},
  \bibinfo{journal}{Phys. Rev. B} \textbf{\bibinfo{volume}{86}},
  \bibinfo{pages}{125119} (\bibinfo{year}{2012}).

\bibitem[{\citenamefont{Qi et~al.}(2008)\citenamefont{Qi, Hughes, and
  Zhang}}]{Qi2008}
\bibinfo{author}{\bibfnamefont{X.-L.} \bibnamefont{Qi}},
  \bibinfo{author}{\bibfnamefont{T.~L.} \bibnamefont{Hughes}},
  \bibnamefont{and} \bibinfo{author}{\bibfnamefont{S.-C.} \bibnamefont{Zhang}},
  \bibinfo{journal}{Phys. Rev. B} \textbf{\bibinfo{volume}{78}},
  \bibinfo{pages}{195424} (\bibinfo{year}{2008}).

\bibitem[{\citenamefont{Qi and Zhang}(2011)}]{Qi2011}
\bibinfo{author}{\bibfnamefont{X.-L.} \bibnamefont{Qi}} \bibnamefont{and}
  \bibinfo{author}{\bibfnamefont{S.-C.} \bibnamefont{Zhang}},
  \bibinfo{journal}{Rev. Mod. Phys.} \textbf{\bibinfo{volume}{83}},
  \bibinfo{pages}{1057} (\bibinfo{year}{2011}).

\bibitem[{\citenamefont{Xiao et~al.}(2010)\citenamefont{Xiao, Chang, and
  Niu}}]{Xiao2010}
\bibinfo{author}{\bibfnamefont{D.}~\bibnamefont{Xiao}},
  \bibinfo{author}{\bibfnamefont{M.-C.} \bibnamefont{Chang}}, \bibnamefont{and}
  \bibinfo{author}{\bibfnamefont{Q.}~\bibnamefont{Niu}}, \bibinfo{journal}{Rev.
  Mod. Phys.} \textbf{\bibinfo{volume}{82}}, \bibinfo{pages}{1959}
  (\bibinfo{year}{2010}).

\bibitem[{\citenamefont{Yip}(2014)}]{Yip2014}
\bibinfo{author}{\bibfnamefont{S.}~\bibnamefont{Yip}}, \bibinfo{journal}{Annu.
  Rev. Condens. Matter Phys.} \textbf{\bibinfo{volume}{5}}, \bibinfo{pages}{15}
  (\bibinfo{year}{2014}).

\bibitem[{\citenamefont{Volovik}(1993)}]{Volovik1993}
\bibinfo{author}{\bibfnamefont{G.~E.} \bibnamefont{Volovik}},
  \bibinfo{journal}{JETP Lett.} \textbf{\bibinfo{volume}{58}},
  \bibinfo{pages}{469} (\bibinfo{year}{1993}).

\bibitem[{\citenamefont{Lee}(1993)}]{Lee1993}
\bibinfo{author}{\bibfnamefont{P.~A.} \bibnamefont{Lee}},
  \bibinfo{journal}{Phys. Rev. Lett.} \textbf{\bibinfo{volume}{71}},
  \bibinfo{pages}{1887} (\bibinfo{year}{1993}).

\bibitem[{\citenamefont{Graf et~al.}(1996)\citenamefont{Graf, Yip, Sauls, and
  Rainer}}]{Graf1996}
\bibinfo{author}{\bibfnamefont{M.~J.} \bibnamefont{Graf}},
  \bibinfo{author}{\bibfnamefont{S.-K.} \bibnamefont{Yip}},
  \bibinfo{author}{\bibfnamefont{J.~A.} \bibnamefont{Sauls}}, \bibnamefont{and}
  \bibinfo{author}{\bibfnamefont{D.}~\bibnamefont{Rainer}},
  \bibinfo{journal}{Phys. Rev. B} \textbf{\bibinfo{volume}{53}},
  \bibinfo{pages}{15147} (\bibinfo{year}{1996}).

\bibitem[{\citenamefont{Qi et~al.}(2010)\citenamefont{Qi, Hughes, and
  Zhang}}]{Qi2010t}
\bibinfo{author}{\bibfnamefont{X.-L.} \bibnamefont{Qi}},
  \bibinfo{author}{\bibfnamefont{T.~L.} \bibnamefont{Hughes}},
  \bibnamefont{and} \bibinfo{author}{\bibfnamefont{S.-C.} \bibnamefont{Zhang}},
  \bibinfo{journal}{Phys. Rev. B} \textbf{\bibinfo{volume}{81}},
  \bibinfo{pages}{134508} (\bibinfo{year}{2010}).

\bibitem[{\citenamefont{Kitaev}(2009)}]{Kitaev2009}
\bibinfo{author}{\bibfnamefont{A.}~\bibnamefont{Kitaev}}, \bibinfo{journal}{AIP
  Conference Proceedings} \textbf{\bibinfo{volume}{1134}}
  (\bibinfo{year}{2009}).

\bibitem[{\citenamefont{Ryu et~al.}(2010)\citenamefont{Ryu, Schnyder, Furusaki,
  and Ludwig}}]{Ryu2010}
\bibinfo{author}{\bibfnamefont{S.}~\bibnamefont{Ryu}},
  \bibinfo{author}{\bibfnamefont{A.~P.} \bibnamefont{Schnyder}},
  \bibinfo{author}{\bibfnamefont{A.}~\bibnamefont{Furusaki}}, \bibnamefont{and}
  \bibinfo{author}{\bibfnamefont{A.~W.~W.} \bibnamefont{Ludwig}},
  \bibinfo{journal}{New J. Phys.} \textbf{\bibinfo{volume}{12}},
  \bibinfo{pages}{065010} (\bibinfo{year}{2010}).

\bibitem[{\citenamefont{Ho\ifmmode~\check{r}\else
  \v{r}\fi{}ava}(2005)}]{Horava2005}
\bibinfo{author}{\bibfnamefont{P.}~\bibnamefont{Ho\ifmmode~\check{r}\else
  \v{r}\fi{}ava}}, \bibinfo{journal}{Phys. Rev. Lett.}
  \textbf{\bibinfo{volume}{95}}, \bibinfo{pages}{016405}
  (\bibinfo{year}{2005}).

\bibitem[{\citenamefont{Zhao and Wang}(2013)}]{Zhao2013}
\bibinfo{author}{\bibfnamefont{Y.~X.} \bibnamefont{Zhao}} \bibnamefont{and}
  \bibinfo{author}{\bibfnamefont{Z.~D.} \bibnamefont{Wang}},
  \bibinfo{journal}{Phys. Rev. Lett.} \textbf{\bibinfo{volume}{110}},
  \bibinfo{pages}{240404} (\bibinfo{year}{2013}).

\bibitem[{\citenamefont{Chiu and Schnyder}(2014)}]{Chiu2014}
\bibinfo{author}{\bibfnamefont{C.-K.} \bibnamefont{Chiu}} \bibnamefont{and}
  \bibinfo{author}{\bibfnamefont{A.~P.} \bibnamefont{Schnyder}},
  \bibinfo{journal}{Phys. Rev. B} \textbf{\bibinfo{volume}{90}},
  \bibinfo{pages}{205136} (\bibinfo{year}{2014}).

\bibitem[{\citenamefont{Schnyder et~al.}(2012)\citenamefont{Schnyder, Brydon,
  and Timm}}]{Schnyder2012}
\bibinfo{author}{\bibfnamefont{A.~P.} \bibnamefont{Schnyder}},
  \bibinfo{author}{\bibfnamefont{P.~M.~R.} \bibnamefont{Brydon}},
  \bibnamefont{and} \bibinfo{author}{\bibfnamefont{C.}~\bibnamefont{Timm}},
  \bibinfo{journal}{Phys. Rev. B} \textbf{\bibinfo{volume}{85}},
  \bibinfo{pages}{024522} (\bibinfo{year}{2012}).

\bibitem[{\citenamefont{Matsuura et~al.}(2013)\citenamefont{Matsuura, Chang,
  Schnyder, and Ryu}}]{Shunji2013}
\bibinfo{author}{\bibfnamefont{S.}~\bibnamefont{Matsuura}},
  \bibinfo{author}{\bibfnamefont{P.-Y.} \bibnamefont{Chang}},
  \bibinfo{author}{\bibfnamefont{A.~P.} \bibnamefont{Schnyder}},
  \bibnamefont{and} \bibinfo{author}{\bibfnamefont{S.}~\bibnamefont{Ryu}},
  \bibinfo{journal}{New J. Phys.} \textbf{\bibinfo{volume}{15}},
  \bibinfo{pages}{065001} (\bibinfo{year}{2013}).

\bibitem[{\citenamefont{Schnyder and Brydon}(2015)}]{Schnyder2015}
\bibinfo{author}{\bibfnamefont{A.~P.} \bibnamefont{Schnyder}} \bibnamefont{and}
  \bibinfo{author}{\bibfnamefont{P.~M.~R.} \bibnamefont{Brydon}},
  \bibinfo{journal}{J. Phys.: Condens. Matter} \textbf{\bibinfo{volume}{27}},
  \bibinfo{pages}{243201} (\bibinfo{year}{2015}).

\bibitem[{\citenamefont{Haldane}(2004)}]{Haldane2004}
\bibinfo{author}{\bibfnamefont{F.~D.~M.} \bibnamefont{Haldane}},
  \bibinfo{journal}{Phys. Rev. Lett.} \textbf{\bibinfo{volume}{93}},
  \bibinfo{pages}{206602} (\bibinfo{year}{2004}).

\bibitem[{\citenamefont{Murakami}(2007)}]{Murakami2007}
\bibinfo{author}{\bibfnamefont{S.}~\bibnamefont{Murakami}},
  \bibinfo{journal}{New J. Phys.} \textbf{\bibinfo{volume}{9}},
  \bibinfo{pages}{356} (\bibinfo{year}{2007}).

\bibitem[{\citenamefont{Hosur et~al.}(2014)\citenamefont{Hosur, Dai, Fang, and
  Qi}}]{Hosur2014}
\bibinfo{author}{\bibfnamefont{P.}~\bibnamefont{Hosur}},
  \bibinfo{author}{\bibfnamefont{X.}~\bibnamefont{Dai}},
  \bibinfo{author}{\bibfnamefont{Z.}~\bibnamefont{Fang}}, \bibnamefont{and}
  \bibinfo{author}{\bibfnamefont{X.-L.} \bibnamefont{Qi}},
  \bibinfo{journal}{Phys. Rev. B} \textbf{\bibinfo{volume}{90}},
  \bibinfo{pages}{045130} (\bibinfo{year}{2014}).

\bibitem[{\citenamefont{Li et~al.}(2010)\citenamefont{Li, Wang, Qi, and
  Zhang}}]{Li2010}
\bibinfo{author}{\bibfnamefont{R.}~\bibnamefont{Li}},
  \bibinfo{author}{\bibfnamefont{J.}~\bibnamefont{Wang}},
  \bibinfo{author}{\bibfnamefont{X.-L.} \bibnamefont{Qi}}, \bibnamefont{and}
  \bibinfo{author}{\bibfnamefont{S.-C.} \bibnamefont{Zhang}},
  \bibinfo{journal}{Nat. Phys.} \textbf{\bibinfo{volume}{6}},
  \bibinfo{pages}{284} (\bibinfo{year}{2010}).

\bibitem[{\citenamefont{Bulmash et~al.}(2015)\citenamefont{Bulmash, Hosur,
  Zhang, and Qi}}]{Bulmash2015}
\bibinfo{author}{\bibfnamefont{D.}~\bibnamefont{Bulmash}},
  \bibinfo{author}{\bibfnamefont{P.}~\bibnamefont{Hosur}},
  \bibinfo{author}{\bibfnamefont{S.-C.} \bibnamefont{Zhang}}, \bibnamefont{and}
  \bibinfo{author}{\bibfnamefont{X.-L.} \bibnamefont{Qi}},
  \bibinfo{journal}{Phys. Rev. X} \textbf{\bibinfo{volume}{5}},
  \bibinfo{pages}{021018} (\bibinfo{year}{2015}).

\bibitem[{\citenamefont{Wang et~al.}(2011)\citenamefont{Wang, Qi, and
  Zhang}}]{Wang2011}
\bibinfo{author}{\bibfnamefont{Z.}~\bibnamefont{Wang}},
  \bibinfo{author}{\bibfnamefont{X.-L.} \bibnamefont{Qi}}, \bibnamefont{and}
  \bibinfo{author}{\bibfnamefont{S.-C.} \bibnamefont{Zhang}},
  \bibinfo{journal}{Phys. Rev. B} \textbf{\bibinfo{volume}{84}},
  \bibinfo{pages}{014527} (\bibinfo{year}{2011}).

\bibitem[{\citenamefont{Furusaki et~al.}(2013)\citenamefont{Furusaki, Nagaosa,
  Nomura, Ryu, and Takayanagi}}]{Furusaki2013}
\bibinfo{author}{\bibfnamefont{A.}~\bibnamefont{Furusaki}},
  \bibinfo{author}{\bibfnamefont{N.}~\bibnamefont{Nagaosa}},
  \bibinfo{author}{\bibfnamefont{K.}~\bibnamefont{Nomura}},
  \bibinfo{author}{\bibfnamefont{S.}~\bibnamefont{Ryu}}, \bibnamefont{and}
  \bibinfo{author}{\bibfnamefont{T.}~\bibnamefont{Takayanagi}},
  \bibinfo{journal}{Comptes Rendus Physique} \textbf{\bibinfo{volume}{14}},
  \bibinfo{pages}{871 } (\bibinfo{year}{2013}), ISSN \bibinfo{issn}{1631-0705}.

\bibitem[{\citenamefont{Witten}(1995)}]{Witten:1992fb}
\bibinfo{author}{\bibfnamefont{E.}~\bibnamefont{Witten}},
  \bibinfo{journal}{Prog. Math.} \textbf{\bibinfo{volume}{133}},
  \bibinfo{pages}{637} (\bibinfo{year}{1995}), \eprint{hep-th/9207094}.

\bibitem[{\citenamefont{Ooguri and Vafa}(2000)}]{Ooguri2000}
\bibinfo{author}{\bibfnamefont{H.}~\bibnamefont{Ooguri}} \bibnamefont{and}
  \bibinfo{author}{\bibfnamefont{C.}~\bibnamefont{Vafa}},
  \bibinfo{journal}{Nucl. Phys. B} \textbf{\bibinfo{volume}{577}},
  \bibinfo{pages}{419 } (\bibinfo{year}{2000}), ISSN \bibinfo{issn}{0550-3213}.

\bibitem[{\citenamefont{Anderson}(1959)}]{Anderson1959}
\bibinfo{author}{\bibfnamefont{P.~W.} \bibnamefont{Anderson}},
  \bibinfo{journal}{J. Phys. Chem. Solids} \textbf{\bibinfo{volume}{11}},
  \bibinfo{pages}{26 } (\bibinfo{year}{1959}).

\bibitem[{trs()}]{trs}
\bibinfo{note}{For simplicity, we assume here $\psi_\mathbf{k}$ is chosen so
  that $\mathcal{T}$ does not depend on $\mathbf{k}$. Choosing another basis
  where $\mathcal{T}$ depends on $\mathbf{k}$ does not change the conclusions
  hereafter.}

\bibitem[{\citenamefont{Tsuei and Kirtley}(2000)}]{Tsuei2000}
\bibinfo{author}{\bibfnamefont{C.~C.} \bibnamefont{Tsuei}} \bibnamefont{and}
  \bibinfo{author}{\bibfnamefont{J.~R.} \bibnamefont{Kirtley}},
  \bibinfo{journal}{Rev. Mod. Phys.} \textbf{\bibinfo{volume}{72}},
  \bibinfo{pages}{969} (\bibinfo{year}{2000}).

\bibitem[{\citenamefont{Wang and Lee}(2011)}]{Wangf2011}
\bibinfo{author}{\bibfnamefont{F.}~\bibnamefont{Wang}} \bibnamefont{and}
  \bibinfo{author}{\bibfnamefont{D.-H.} \bibnamefont{Lee}},
  \bibinfo{journal}{Science} \textbf{\bibinfo{volume}{332}},
  \bibinfo{pages}{200} (\bibinfo{year}{2011}).

\bibitem[{\citenamefont{Berry}(1984)}]{Berry1984}
\bibinfo{author}{\bibfnamefont{M.~V.} \bibnamefont{Berry}},
  \bibinfo{journal}{Proc. R. Soc. Lond. A} \textbf{\bibinfo{volume}{392}},
  \bibinfo{pages}{45} (\bibinfo{year}{1984}), ISSN \bibinfo{issn}{0080-4630}.

\bibitem[{\citenamefont{Ryu et~al.}(2012)\citenamefont{Ryu, Moore, and
  Ludwig}}]{Ryu2012}
\bibinfo{author}{\bibfnamefont{S.}~\bibnamefont{Ryu}},
  \bibinfo{author}{\bibfnamefont{J.~E.} \bibnamefont{Moore}}, \bibnamefont{and}
  \bibinfo{author}{\bibfnamefont{A.~W.~W.} \bibnamefont{Ludwig}},
  \bibinfo{journal}{Phys. Rev. B} \textbf{\bibinfo{volume}{85}},
  \bibinfo{pages}{045104} (\bibinfo{year}{2012}).

\bibitem[{\citenamefont{Forward}(1961)}]{Forward1961}
\bibinfo{author}{\bibfnamefont{R.~L.} \bibnamefont{Forward}},
  \bibinfo{journal}{Proceedings of the IRE} \textbf{\bibinfo{volume}{49}},
  \bibinfo{pages}{892} (\bibinfo{year}{1961}), ISSN \bibinfo{issn}{0096-8390}.

\bibitem[{\citenamefont{Clark and Tucker}(2000)}]{Clark2000}
\bibinfo{author}{\bibfnamefont{S.~J.} \bibnamefont{Clark}} \bibnamefont{and}
  \bibinfo{author}{\bibfnamefont{R.~W.} \bibnamefont{Tucker}},
  \bibinfo{journal}{Classical and Quantum Gravity}
  \textbf{\bibinfo{volume}{17}}, \bibinfo{pages}{4125} (\bibinfo{year}{2000}).

\bibitem[{\citenamefont{Wang et~al.}(2016)\citenamefont{Wang, Lian, and
  Zhang}}]{Wang2016}
\bibinfo{author}{\bibfnamefont{J.}~\bibnamefont{Wang}},
  \bibinfo{author}{\bibfnamefont{B.}~\bibnamefont{Lian}}, \bibnamefont{and}
  \bibinfo{author}{\bibfnamefont{S.-C.} \bibnamefont{Zhang}},
  \bibinfo{journal}{Phys. Rev. B} \textbf{\bibinfo{volume}{93}},
  \bibinfo{pages}{045115} (\bibinfo{year}{2016}).

\bibitem[{\citenamefont{Luttinger}(1964)}]{Luttinger1964}
\bibinfo{author}{\bibfnamefont{J.~M.} \bibnamefont{Luttinger}},
  \bibinfo{journal}{Phys. Rev.} \textbf{\bibinfo{volume}{135}},
  \bibinfo{pages}{A1505} (\bibinfo{year}{1964}).

\bibitem[{mas()}]{mass}
\bibinfo{note}{The mass term $m$ can have zeros on a nodal line, but the
  superconductor will have gapless nodal points and the physical meanings
  $\theta$ and $S_\theta$ are still ambiguous.}

\bibitem[{\citenamefont{Witten}(1989)}]{Witten1989}
\bibinfo{author}{\bibfnamefont{E.}~\bibnamefont{Witten}},
  \bibinfo{journal}{Commun. Math. Phys.} \textbf{\bibinfo{volume}{121}},
  \bibinfo{pages}{351} (\bibinfo{year}{1989}), ISSN \bibinfo{issn}{1432-0916}.

\bibitem[{\citenamefont{Polyakov}(1988)}]{Polyakov1988}
\bibinfo{author}{\bibfnamefont{A.~M.} \bibnamefont{Polyakov}},
  \bibinfo{journal}{Mod. Phys. Lett. A} \textbf{\bibinfo{volume}{03}},
  \bibinfo{pages}{325} (\bibinfo{year}{1988}).

\bibitem[{ske()}]{skein}
\bibinfo{note}{For U($1$) Chern-Simons theories, the skein relation does not
  define a knot polynomial since $r=0$.}

\bibitem[{fra()}]{framing}
\bibinfo{note}{We can try to define a canonical definition of framing of the
  nodal line by how $d\lambda$ wraps around as we go along the nodal line.
  However, one can argue that no such local definition of framing would be
  compatible with our results: When two segments of nodal lines cross,
  regardless of whether they are part of the same nodal line or not, the
  framing numbers would change by the same amount. However, the linking
  contribution should change only if they are part of different nodal lines, in
  contradiction with our local computation which is insensitive to the global
  properties of the nodal lines.}

\bibitem[{\citenamefont{Wilczek and Zee}(1984)}]{Wilczek1984}
\bibinfo{author}{\bibfnamefont{F.}~\bibnamefont{Wilczek}} \bibnamefont{and}
  \bibinfo{author}{\bibfnamefont{A.}~\bibnamefont{Zee}},
  \bibinfo{journal}{Phys. Rev. Lett.} \textbf{\bibinfo{volume}{52}},
  \bibinfo{pages}{2111} (\bibinfo{year}{1984}).

\bibitem[{gen()}]{gene}
\bibinfo{note}{Such a choice of $T^0$ is most natural in the sense that one
  gets a normalized trace relation $\mbox{Tr}(T^aT^b)=\delta^{ab}/2$.}

\bibitem[{\citenamefont{Ran et~al.}(2009)\citenamefont{Ran, Zhang, and
  Vishwanath}}]{Ran2009}
\bibinfo{author}{\bibfnamefont{Y.}~\bibnamefont{Ran}},
  \bibinfo{author}{\bibfnamefont{Y.}~\bibnamefont{Zhang}}, \bibnamefont{and}
  \bibinfo{author}{\bibfnamefont{A.}~\bibnamefont{Vishwanath}},
  \bibinfo{journal}{Nat. Phys.} \textbf{\bibinfo{volume}{5}},
  \bibinfo{pages}{298} (\bibinfo{year}{2009}).

\bibitem[{\citenamefont{Slager et~al.}(2013)\citenamefont{Slager, Mesaros,
  Juricic, and Zaanen}}]{Slager2013}
\bibinfo{author}{\bibfnamefont{R.-J.} \bibnamefont{Slager}},
  \bibinfo{author}{\bibfnamefont{A.}~\bibnamefont{Mesaros}},
  \bibinfo{author}{\bibfnamefont{V.}~\bibnamefont{Juricic}}, \bibnamefont{and}
  \bibinfo{author}{\bibfnamefont{J.}~\bibnamefont{Zaanen}},
  \bibinfo{journal}{Nat. Phys.} \textbf{\bibinfo{volume}{9}},
  \bibinfo{pages}{98} (\bibinfo{year}{2013}).

\bibitem[{\citenamefont{Slager et~al.}(2014)\citenamefont{Slager, Mesaros,
  Juri\ifmmode \check{c}\else \v{c}\fi{}i\ifmmode~\acute{c}\else \'{c}\fi{},
  and Zaanen}}]{Slager2014}
\bibinfo{author}{\bibfnamefont{R.-J.} \bibnamefont{Slager}},
  \bibinfo{author}{\bibfnamefont{A.}~\bibnamefont{Mesaros}},
  \bibinfo{author}{\bibfnamefont{V.}~\bibnamefont{Juri\ifmmode \check{c}\else
  \v{c}\fi{}i\ifmmode~\acute{c}\else \'{c}\fi{}}}, \bibnamefont{and}
  \bibinfo{author}{\bibfnamefont{J.}~\bibnamefont{Zaanen}},
  \bibinfo{journal}{Phys. Rev. B} \textbf{\bibinfo{volume}{90}},
  \bibinfo{pages}{241403} (\bibinfo{year}{2014}),
  \urlprefix\url{http://link.aps.org/doi/10.1103/PhysRevB.90.241403}.

\bibitem[{Lsp()}]{Lspace}
\bibinfo{note}{A Lagrangian space in the phase space $(x,y,z,k_x,k_y,k_z)$
  (which is symplectic) is a submanifold where the symplectic form vanishes.}

\bibitem[{\citenamefont{Freed and Moore}(2013)}]{Freed2013}
\bibinfo{author}{\bibfnamefont{D.~S.} \bibnamefont{Freed}} \bibnamefont{and}
  \bibinfo{author}{\bibfnamefont{G.~W.} \bibnamefont{Moore}},
  \bibinfo{journal}{Annales Henri Poincar{\'e}} \textbf{\bibinfo{volume}{14}},
  \bibinfo{pages}{1927} (\bibinfo{year}{2013}), ISSN \bibinfo{issn}{1424-0661}.

\bibitem[{\citenamefont{Kohmoto et~al.}(1992)\citenamefont{Kohmoto, Halperin,
  and Wu}}]{Kohmoto1992}
\bibinfo{author}{\bibfnamefont{M.}~\bibnamefont{Kohmoto}},
  \bibinfo{author}{\bibfnamefont{B.~I.} \bibnamefont{Halperin}},
  \bibnamefont{and} \bibinfo{author}{\bibfnamefont{Y.-S.} \bibnamefont{Wu}},
  \bibinfo{journal}{Phys. Rev. B} \textbf{\bibinfo{volume}{45}},
  \bibinfo{pages}{13488} (\bibinfo{year}{1992}).

\bibitem[{\citenamefont{Lian and Zhang}(2016)}]{Lian2016}
\bibinfo{author}{\bibfnamefont{B.}~\bibnamefont{Lian}} \bibnamefont{and}
  \bibinfo{author}{\bibfnamefont{S.-C.} \bibnamefont{Zhang}},
  \bibinfo{journal}{Phys. Rev. B} \textbf{\bibinfo{volume}{94}},
  \bibinfo{pages}{041105} (\bibinfo{year}{2016}).

\end{thebibliography}

\end{document}